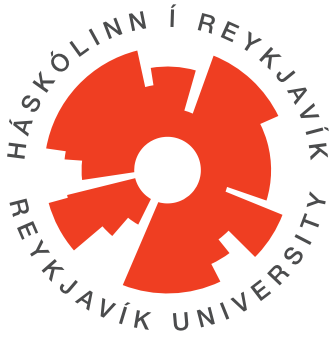

Technical Report

# A Choreographed Outline Instrumentation Approach for Asynchronous Components

Luca Aceto, Duncan Paul Attard, Adrian Francalanza, and Anna Ingólfsdóttir

**April, 2021**

**TÖLVUNARFRÆÐIDEILD**
DEPARTMENT OF COMPUTER SCIENCE

# A Choreographed Outline Instrumentation Algorithm for Asynchronous Components

LUCA ACETO, Reykjavík University, Iceland and Gran Sasso Science Institute, Italy
DUNCAN PAUL ATTARD, University of Malta, Malta and Reykjavík University, Iceland
ADRIAN FRANCALANZA, University of Malta, Malta
ANNA INGÓLFSDÓTTIR, Reykjavík University, Iceland

The runtime analysis of decentralised software requires instrumentation methods that are scalable, but also minimally invasive. This paper presents a new algorithm that instruments choreographed outline monitors. Our instrumentation algorithm scales and reorganises monitors dynamically as the system executes. We demonstrate the implementability of choreographed outline instrumentation and compare it to inline instrumentation, subject to rigorous and comprehensive benchmarking. Our results debunk the general notion that outline monitoring is necessarily infeasible, and show that our implementation induces runtime overhead comparable to that of its inline counterpart for many practical cases.

## 1 INTRODUCTION

Software is increasingly decentralised and dynamic, in both structure and scale. Constituent system components execute concurrently: they are dynamically created and destroyed, forging interconnections that are determined at runtime. These characteristics make the expected software behaviour, in terms of both correctness and performance, harder to establish. Runtime monitoring is a widespread technique that is used for analysing the behaviour of decentralised and distributed software *after* it has been deployed [43, 44]. The technique uses *monitors*—computational entities consisting of logically-distinct *instrumentation* and *analysis* units—to observe the execution of the System under Scrutiny (SuS). *Instrumentation* lies at the heart of runtime monitoring [11, 43, 56]. It refers to the extraction of information from executing software, following one of two approaches. In the *grey-box* approach, developers implement instrumentation by *weaving* tracing instructions into the SuS. Alternatively, instrumentation can leverage *tracing infrastructures* (*e.g.* DTrace [20], LTTng [34], OpenJ9 Trace [35]) to collect trace events externally, thereby treating the SuS as a *black-box*.

Runtime Verification (RV) [11] is the verification complement to monitoring. Monitors are used to process the trace events extracted from the executing SuS in order to determine whether a *correctness specification* is satisfied or violated [11, 43]. A wide variety of RV tools instrument the SuS by weaving the monitor logic to maintain low runtime overhead [38, 39]. This grey-box approach, known as *monitor inlining* [40], brings with it other benefits such as timely detections [11]. However, inlining is not necessarily the best approach for the RV of large-scale decentralised open systems [44]. For cases where the SuS sources or binaries are unavailable, *e.g.* closed-source components, licensing agreements, third-party services, *etc.*, inlining *cannot* be used. In most cases, inlining is programming-language dependent, making it difficult to apply to heterogeneous components. Additionally, inlining is hard to undo once the instrumented SuS is in place.

*Outline monitoring* [40] is an alternative approach to inlining where the SuS and monitor components are organised as *concurrent* logical entities. This *minimal coupling* between the SuS and monitors begets a number of advantages that are more attuned to the characteristics of decentralised systems. In particular, monitors can be enabled or disabled at runtime *without* system redeployments or restarts. This is invaluable when conducting live debugging to track subtle concurrency issues that only emerge for certain execution paths. Outlining does not necessitate access to the SuS sources and tends to be more language-agnostic. Moreover, the independent SuS and monitor components permit a degree of *partial failure*: a faulty monitor does not compromise the running system, and reciprocally, a crashed SuS component can still be detected by the external monitor.

Despite these merits, inlining is still the predominant method employed by tools targeting decentralised RV (*e.g.* Basin et al. [12], Bonakdarpour et al. [18], Cassar and Francalanza [21], Colombo et al. [32], El-Hokayem and Falcone [36, 37], Fraigniaud et al. [42], Jin et al. [51], Kim et al. [54], Sen et al. [77, 78]). One probable reason for this is *legacy tooling*. A number of decentralised RV tools are extensions of older efforts that were originally conceived for monolithic, single-threaded RV, where inlining has traditionally performed well. It is therefore advantageous to extend a tried-and-tested approach to a new domain such as decentralised systems, rather than abandoning this technological investment in favour of a completely new approach.

The second reason is that concurrent systems have more stringent requirements, making decentralised outline instrumentation *harder* to build. Decentralised software is often designed to *scale* in response to fluctuating demands, and this naturally requires a RV set-up that grows and shrinks accordingly. With inlining, scalability comes about as a byproduct of the monitor logic that is weaved into components of the SuS. By contrast, scaling must be *explicitly* engineered in the case of outline monitoring, since the instrumentation needs to *dynamically* reconfigure the decentralised RV set-up while the runtime analysis is being effected. The decoupling forces instrumentation to contend with the *race conditions* (*e.g.* message reordering) that arise from the asynchronous execution of the SuS and monitors. In particular, decentralised outline instrumentation should ensure that the trace events collected from the SuS reach the intended monitors in the *correct order*. To our knowledge, no outline instrumentation that is decentralised and scalable currently exists.

The third reason why outline instrumentation is seldom considered for decentralised RV is that outlining is *perceived* to induce infeasibly higher runtime overhead when compared to inlining. This perception partly stems from the fact that inlining identifies the designated monitor instrumentation points in the SuS prior to deployment, whereas outlining defers this decision *post-deployment*. Moreover, overhead for decentralised RV is still measured in terms of criteria applicable to monolithic, single-threaded systems such as total execution time (*e.g.* see Bodden et al. [16], Chen and Rosu [25, 26], Lange and Yoshida [55], Neykova and Yoshida [64, 65], Reger et al. [70]). In distributed settings with high latencies, there are other facets of runtime overhead that are as important (if not more), such as the mean response time of a service invocation.

*Contributions.* Monitor decentralisation and outlining necessitate a *choreographed instrumentation* approach—here called 'choreographed outline instrumentation', and referred to simply as 'outline instrumentation' in the sequel. Our contributions target the last two reasons inhibiting the adoption of outline monitoring outlined above, *i.e.,* the difficulty of building outline decentralised instrumentation and the perception of unfeasibly high overheads:

  (i) We devise the first general algorithm for instrumenting choreographed outline monitors that scales with the SuS. We discuss the technical challenges of outlining, elaborating on the issues that arise due to the dynamic reconfiguration of choreographed monitors in an asynchronous setting, Sec. 2;

  (ii) We demonstrate the implementability of our proposed choreographed outline monitoring algorithm by building a tool using an industry-strength programming concurrent language. We validate the correctness of our implementation by subjecting it to rigorous invariance testing, Sec. 3.
  (iii) We compare implementations of inline and choreographed outline instrumentation via a comprehensive empirical study and show that, for respectable loads, both approaches induce comparable overhead. We judge this performance on multiple facets of runtime overhead, subjecting the SuS to loads that are typically observed in practice [3]. Our estimated overhead per SuS component reveals that outline instrumentation exacts negligible penalties, Sec. 4.

## 2 CHOREOGRAPHED OUTLINE INSTRUMENTATION

Studies on monolithic RV conflate the instrumentation of monitors *and* the analysis of trace events. Generally, extracting trace events via instrumentation induces the larger part of the overhead, when compared to the analysis [46, 63]. APM tools often resort to *sampling* as means of managing the overhead induced by the instrumentation [47, 80]. This solution is, however, inapplicable to RV, since many verification applications require that *every* event is considered in strict orderly fashion. In our study of decentralised RV, we delineate between the instrumentation and analysis to: *(i)* isolate and address the complications of instrumenting decentralised outline monitors, and, *(ii)* understand the impact of separating the instrumentation and analysis w.r.t. overhead. The outline instrumentation algorithm makes the following minimal assumptions on our operational model for concurrent components:

  $A_1$ *Local clocks.* Components do not share a common global clock.
  $A_2$ *Dynamic system.* The number of components fluctuates.
  $A_3$ *Point-to-point communication.* A sender component can interact with one receiver at any point in time.
  $A_4$ *Reliable communication.* Messages sent are not tampered with, are guaranteed to be delivered, and are never duplicated.
  $A_5$ *Message reordering.* The order the messages are sent is not necessarily the order received in. This does not apply to point-to-point communication, *i.e.,* successive messages exchanged between component pairs are delivered in the same sequence issued.
  $A_6$ *Reliable components.* Components are not subject to either fail-stops or Byzantine failures.

Our outline instrumentation requires monitors to be:

  $R_1$ *Passive.* Monitors only react to SuS events, and do not interfere with its execution.
  $R_2$ *Decentralised.* No central entity coordinates monitors.
  $R_3$ *Reliable.* Trace events are not lost, nor analysed out-of-order.

We use sequence recognisers [57, 74] as RV monitors that reach irrevocable verdicts after analysing a *finite stream* of trace events [2]; they satisfy requirement $R_1$. Our monitors are instrumented to run asynchronously with the SuS, in line with assumption $A_1$; this turns out to be the general case for distributed set-ups. We remark that distribution can be obtained by weakening assumptions $A_4$ and $A_6$. Assumption $A_2$ and requirement $R_2$ call for monitoring to scale dynamically by continually reconfiguring the choreography in response to specific events exhibited by the SuS. Requirement $R_3$ guards against issues arising from assumption $A_5$—this is vital for analyses that are sensitive to the temporal ordering of trace events (*e.g.* RV, live debugging).

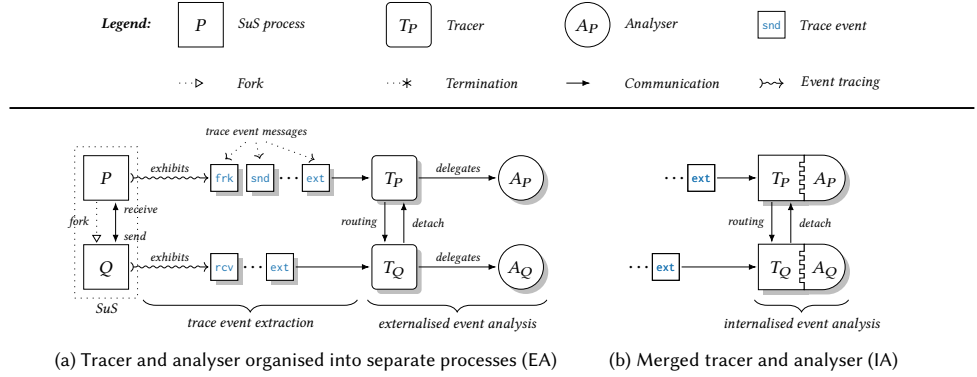

(a) Tracer and analyser organised into separate processes (EA)  (b) Merged tracer and analyser (IA)

Fig. 1. Outline verdict monitoring set-up consisting of tracer and analyser roles

## 2.1 Overview

Fig. 1 shows the variants of outline instrumentation we consider. The externalised analysis (EA) arrangement in fig. 1a consists of independently-executing *tracer* and *analyser* processes, teasing apart the tasks of trace event handing and monitor reorganisation, performed by tracers, from the task of trace event examination, effected by analysers. Separating these roles follows the single responsibility principle advocated in component-based design [4, 58] at the expense of introducing an additional analyser component. By contrast, the internalised analysis (IA) variant merges the tracing and analysis aspects to forgo this extra component, as in fig. 1b. Our algorithm works for both grey-box and black-box tracing, so long as it is provided with streams of trace event *messages* for the monitored SuS components. Tracers may start or stop these event streams at runtime. In fig. 1, the trace event messages from processes $P$ and $Q$ are directed to the respective tracers $T_P$ and $T_Q$, where they are analysed by $A_P$ and $A_Q$. Our model assumes that:

- $A_7$ *Tracers cannot share system processes.* A system process can be traced by one tracer at any point in time.
- $A_8$ *System processes may share tracers.* A tracer may trace multiple system processes concurrently.
- $A_9$ *System processes inherit tracers.* A forked system process is automatically assigned the tracer of its parent process.

Assumption $A_7$ means that for a tracer to start tracing some system process being traced, it must first *stop* the active tracer before it can take over and *continue* tracing said process itself.

## 2.2 The System Model

*Processes.* In concurrency, processes are created *hierarchically*, starting from the top-level process that *forks* other processes, and so forth, *e.g.* `CreateThread()` in Windows [61], `pthread_create()` for POSIX threads [19], `ActorContext.spawn()` in Akka [71], `spawn()` in Erlang [24]. We borrow standard terminology (root, child, ancestor, *etc.*) to describe the relationships between processes in the tree. Our model assumes a denumerable set of process identifiers (PIDs) to refer to processes. We distinguish between system, tracer and analyser process forms, denoting them respectively by the sets $\text{PID}_s$, $\text{PID}_t$ and $\text{PID}_a$, where $p_s \in \text{PID}_s$, $p_t \in \text{PID}_t$, $p_a \in \text{PID}_a$. New processes are created via the function $\text{fork}(g)$ that takes the signature of the code to be run by the forked process, $g \in \text{SIG}$, returning its *fresh* PID. We refer to the process invoking fork as the *parent*, and to the forked process as the *child*. To create analyser processes, the function fork is overloaded to accept *verdict-flagging* code, $v \in \text{MON}$, and return the corresponding PID $p_a$; tracer processes are forked analogously.

A Choreographed Outline Instrumentation Algorithm for Asynchronous Components

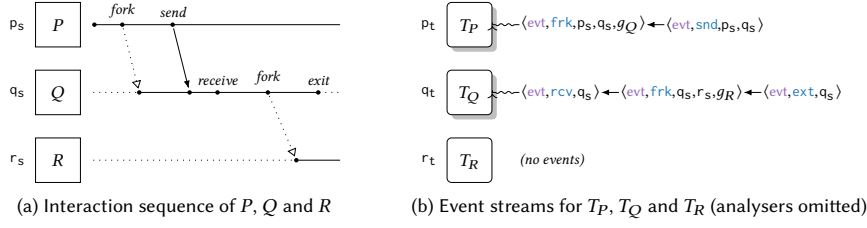

(a) Interaction sequence of $P$, $Q$ and $R$     (b) Event streams for $T_P$, $T_Q$ and $T_R$ (analysers omitted)

Fig. 2. SuS with processes $P$, $Q$, and $R$ instrumented with three independent monitors

Processes communicate through asynchronous messages. Each process owns a *message queue*, K, from where it can read messages *out-of-order* and in a *non-blocking* fashion. We use the terms *tracer* and *monitor* synonymously wherever the distinction is unimportant; the term *component* refers to a grouping of one or more processes.

*Messages.* We represent messages, $m \in \text{Msg}$, as tuples $\langle q, d_0, d_1, \ldots, d_n \rangle$, where $q \in \{\text{evt}, \text{dtc}, \text{rtd}\}$ is the *qualifier* indicating the message type, and $d_{i \in \mathbb{Z}^{0+}}$ are the data elements comprising the payload. The qualifier designates messages for *one* of these purposes:

$q = \text{evt}$ : analysable system *trace events* obtained via tracing,
$q = \text{dtc}$ : *detach commands* that tracers exchange to reorganise the monitoring choreography, or,
$q = \text{rtd}$ : *embedded* evt or dtc messages that are *routed* between tracers.

The meta-variables $e$, $c$, and $r$ are reserved for messages of types evt, dtc and rtd respectively; $m$, refers to generic messages otherwise. We use the dot notation (.) to access specific message data elements through indexable *field names*, e.g. the message qualifier $q$ is read through $m.\text{type}$. Trace events are encoded as messages, $\langle q = \text{evt}, d_0 = a, d_1, \ldots, d_n \rangle$, where $a \in \text{Act}$ identifies the kind of action exhibited by the SuS and $d_1, \ldots, d_n$ designate the data particular to the event. The event action is accessed using $e.\text{act}$. Let $\text{Act} = \{\text{frk}, \text{ext}, \text{snd}, \text{rcv}\}$ denote the process actions fork (frk), exit (ext), send '!' (snd) and receive (rcv). We abuse the notation and use the action name $a$ in lieu of the full trace event message payload (*i.e.*, omitting $q$ and $d_1, \ldots, d_n$) when this simplifies our explanation. The events and corresponding data carried are catalogued in tbl. 1.

| Event | Action ($e$.act) | Field names | Description |
|---|---|---|---|
| fork | frk | src | PID $p_s$ of the (parent) process invoking fork($g$) |
|  |  | tgt | PID $p_s$ of the forked (child) process |
|  |  | sig | Code signature $g$ run by the forked process |
| exit | ext | src | PID $p_s$ of the terminated process |
| send | snd | src | PID $p_s$ of the sender process |
|  |  | tgt | PID $p_s$ of the recipient process |
| receive | rcv | src | PID $p_s$ of the recipient process |

Tbl. 1. Trace event messages data field names

### 2.3 The Running Example

Our SuS consists of three processes where $P$ forks $Q$ and communicates with it; afterwards $Q$ forks $R$ and terminates. $P$, $Q$, and $R$ are assigned PIDs, $p_s$, $q_s$, and $r_s$ respectively. This interaction, captured in fig. 2a, is essentially sequential due to the synchronous dependency between processes; for instance, $Q$ comes into existence after $P$ forks it, and $R$ is forked by $Q$ only when $P$ sends a message which $Q$ receives. We set our SuS with independent monitors, one for each of $P$, $Q$, and $R$, denoting them by $T_P$, $T_Q$ and $T_R$. Fig. 2b shows these monitors, corresponding PIDs, and the *correct* sequence of events each tracer is meant to observe.

Despite its modest size and sequential operation, our SuS and outline monitoring set-up may still be subject to multiple interleaved executions. These result from the asynchronous organisation of the SuS and monitor components, whose execution is at the mercy of external factors such as process scheduling, network latency, *etc*. Outline monitoring must handle all possible component interleaving to *guarantee* that every trace event is reported in the *correct order* reflecting the true behaviour of the SuS, regardless of potential race conditions that arise as a product of non-determinism. Monitor inlining is spared these complications, since static weaving necessarily fixes the precise order in which instrumented code instructions execute. Our algorithm given next details how choreographed outline instrumentation is able to ensure the correct order of trace events, and how scalable monitoring is achieved via dynamic instrumentation and monitor garbage collection.

### 2.4 The Choreographed Instrumentation Approach

The algorithm covers the two variants of fig. 1. Lsts. 1–3 describe the core logic found in each tracer; auxiliary logic is elided due to space limitations. In the pseudocode, we highlight whether the tracer dispatches trace events to the analyser (EA), or analyses these internally (IA)—regardless, tracing is *agnostic* of the analyser logic. Each tracer maintains an internal state $\sigma$ that consists of three maps: *(i)* the *routing map*, $\Pi$, governing how events are routed to other tracers, *(ii)* the *instrumentation map*, $\Phi$, dictating whether new tracers need to be launched, and, *(iii)* the *traced-processes map*, $\Gamma$, recording the system processes that the tracer currently tracks.

*Tracing.* The operations TRACE, CLEAR and PREEMPT provide access to our tracing infrastructure. TRACE enables a tracer with PID $p_t$ to register its interest in receiving trace event messages of a system process with PID $p_s$. This operation can be undone using CLEAR, which *blocks* the calling tracer $p_t$ and returns once all the event messages for $p_s$ that are in transit to $p_t$ have been delivered. PREEMPT combines CLEAR and TRACE, enabling a different tracer $p'_t$ to take over the tracing of process $p_s$ from the current tracer, $p_t$. Following assumption $A_9$, tracing is *inherited* by every child process that a traced system process forks; CLEAR or PREEMPT can be used to alter this arrangement (see lst. 6 in app. A).

*Decentralised trace processing.* Tracers are programmed to react to frk and ext events in the trace. Fig. 3 illustrates how the process creation sequence of the SuS is exploited to instrument tracers. A tracer instruments other tracers whenever it encounters frk events. In fig. 3a, the root tracer $T_P$ analyses the top-level process $P$, step ①. It instruments a new tracer, $T_Q$, for process $Q$ when it observes the fork event $\langle \text{evt},\text{frk},p_s,q_s,g_Q \rangle$ exhibited by $P$ in step ③. The field $e$.tgt carried by frk designates the SuS process to be instrumented with the new tracer, $q_s$ in this case. At this point, $T_Q$ *takes over* the tracing of process $Q$ from $T_P$ by invoking PREEMPT to trace $Q$ *independently* of $T_P$, steps ④ and ⑤ in fig. 3b. $T_P$ resumes its analysis, receiving the send event $\langle \text{evt},\text{snd},p_s,q_s \rangle$ in step ⑩ after $P$ messages $Q$ in step ⑥ of fig. 3c. Subsequent frk events observed by $T_P$ and $T_Q$ are handled as described earlier. Figs. 3c and 3d show how the final tracer, $T_R$, is instrumented as $Q$ forks its child $R$. Following assumption $A_9$, we recall that prior to instrumenting

A Choreographed Outline Instrumentation Algorithm for Asynchronous Components

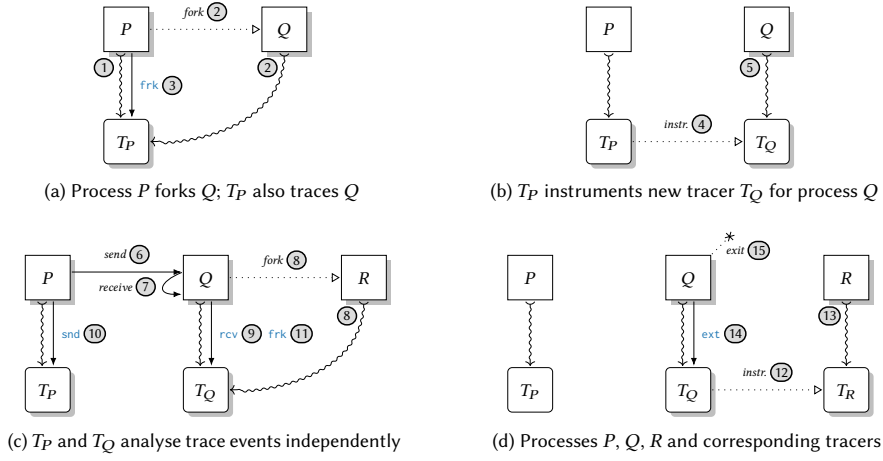

Fig. 3. Outline tracer instrumentation for processes $P$, $Q$ and $Q$ (analysers omitted)

$T_Q$ in step ④, process $Q$ automatically inherits tracer $T_P$ of its parent $P$ in step ②. $T_Q$ is analogously assigned to process $R$ in step ⑧ before $T_Q$ instruments the new tracer $T_R$ for $R$ in step ⑫.

*Trace event routing.* The asynchrony between the SuS and tracer components may give rise to different interleaved executions. Fig. 4 shows one interleaving *alternative* to that of figs. 3b–3d. In fig. 4a, $T_P$ is slow to handle the fork event of $Q$ (received in step ③ in fig. 3a), and fails to instrument $T_Q$ promptly. As a result, the events rcv and frk exhibited by $Q$ are received by $T_P$ in steps ⑦ and ⑨. Fig. 4a shows the case where $\langle \text{evt}, \text{rcv}, q_s \rangle$ is processed by $T_P$, step ⑪, rather than by the *correct* tracer $T_Q$ that would be eventually instrumented by $T_P$. This non-determinism corrupts the runtime analysis, as the events that are to be processed by one tracer reach unwanted tracers.

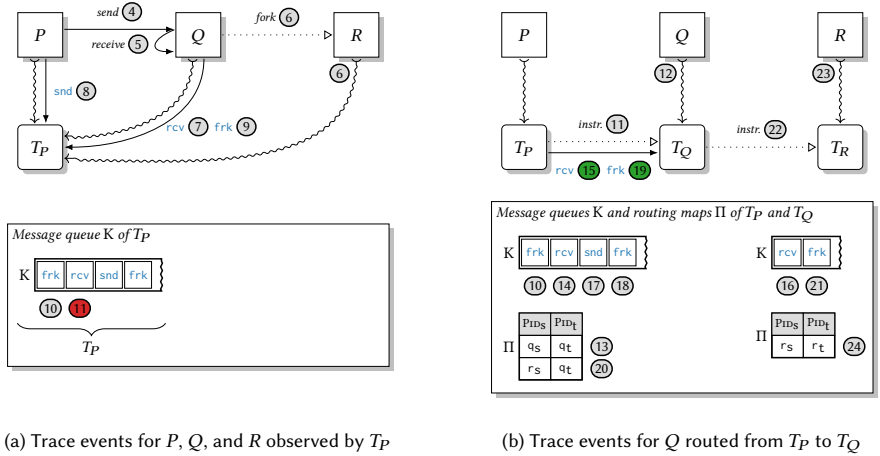

Fig. 4. Hop-by-hop trace event routing using tracer routing maps $\Pi$ (analysers omitted)

To address such issues, tracers *keep* the events they should analyse and *route* the rest to other tracers. This scheme is outlined in fig. 4b, where $T_P$ instruments $T_Q$ with $Q$ in step ⑪. It then processes events ⟨evt,rcv,$q_s$⟩ and ⟨evt,frk,$q_s$,$r_s$,$g_R$⟩, steps ⑭ and ⑱, and *forwards* them to $T_Q$ in steps ⑮ and ⑲. The event ⟨evt,snd,$p_s$,$q_s$⟩ is processed by $T_P$, step ⑰. Concurrently, $T_Q$ acts on the forwarded events in steps ⑯ and ㉑, and $T_R$ is instrumented with $R$ as a result.

Tracers determine which events to keep or forward by means of the *routing map*, $\Pi: \text{PID}_s \rightharpoonup \text{PID}_t$, relating system and tracer PIDs. Each tracer queries its routing map for every event $e$ it processes using the source PID, $e.\text{src}$. An event is forwarded to a tracer with PID $p_t$ when $\Pi(e.\text{src}) = p_t$, otherwise it is handled by the tracer itself since a route for $e.\text{src}$ does not exist, *i.e.*, $\Pi(e.\text{src}) = \bot$. HandleFork, HandleExit and HandleComm in lst. 2 implement this forwarding logic on lines 24, 34 and 44.

A tracer populates its routing map $\Pi$ whenever it processes a fork event ⟨evt,frk,$p_s$,$p'_s$,$g$⟩. It considers *one* of two cases for $p_s$:

$C_1$  $\Pi(p_s) = \bot$. This means that the tracer must adapt the choreography to account for the newly-forked process $p'_s$. It forks a child tracer $T_{P'}$ with PID $p'_t$ to be instrumented with $p'_s$. The mapping $p'_s \mapsto p'_t$ is added to $\Pi$, or,

$C_2$  $\Pi(p_s) = p'_t$. A route to the tracer with PID $p'_t$ exists for trace events originating from $p_s$. The tracer forwards the fork event for process $p_s$ to $p'_t$, and adds the mapping $p'_s \mapsto p'_t$ to $\Pi$.

In cases $C_1$ and $C_2$, the addition of $p'_s \mapsto p'_t$ ensures that future events originating from $p'_s$ can be forwarded onwards. Fig. 4b shows the routing maps of $T_P$ and $T_Q$. $T_P$ adds $q_s \mapsto q_t$, step ⑬, after processing ⟨evt,frk,$p_s$,$q_s$,$g_Q$⟩ from the message queue in step ⑩ and instrumenting tracer $T_Q$ with $Q$ in step ⑪; an instance of case $C_1$. The function Instrument in lst. 1 details this on line 5, where the mapping $e.\text{tgt} \mapsto p'_t$ (with $e.\text{tgt} = p'_s$) is added to $\Pi$, following the creation of tracer $p'_t$. Step ⑳ of fig. 4b constitutes an instance of case $C_2$: $T_P$ adds $r_s \mapsto q_t$ after processing ⟨evt,frk,$q_s$,$r_s$,$g_R$⟩ for $R$ in step ⑱. Crucially, $T_P$ *does not* instrument a new tracer, but delegates this task to $T_Q$ by forwarding the fork event in question. Lines 26 and 81 in lst. 2 (and later line 26 in lst. 3) are manifestations of this, where the mapping $e.\text{tgt} \mapsto p'_t$ is added after the fork event $e$ is routed to the next tracer $p'_t$.

Note that in fig. 4b, the mappings inside $T_P$ point to tracer $T_Q$, and the mapping in $T_Q$ points to $T_R$. This arises as a result of cases $C_1$ and $C_2$, where any tracer in the monitoring choreography can *only* forward events to *adjacent* tracers. For instance, events that $R$ might exhibit and that are collected by $T_P$ must be forwarded twice to reach the intended tracer $T_R$: from tracer $T_P$ to $T_Q$, and from $T_Q$ to $T_R$. This *hop-by-hop routing* [9] between tracers forms a connected DAG,

```
Expect: e.act = frk
 1  def Instrument∘(σ,e,pt)
 2    ps ← e.tgt
 3    if (v ← σ.Φ(e.sig)) ≠ ⊥ then
 4      p't ← fork(Tracer(σ,v,ps,pt))
 5      σ.Π ← σ.Π ∪ {⟨ps,p't⟩}
 6    else
        # In ∘ mode, there is no process ps to detach from
        # a router tracer; add ps to Γ in ∘ mode
 7      σ.Γ ← σ.Γ ∪ {⟨ps,∘⟩}
 8    end if
 9    return σ
10  end def
```

```
Expect: e.act = frk
11  def Instrument•(σ,e,pt)
12    ps ← e.tgt
13    if (v ← σ.Φ(e.sig)) ≠ ⊥ then
14      p't ← fork(Tracer(σ,v,ps,pt))
15      σ.Π ← σ.Π ∪ {⟨ps,p't⟩}
16    else
        # Detach process ps from router tracer pt
17      Detach(ps,pt)
18      σ.Γ ← σ.Γ ∪ {⟨ps,•⟩}  # Add ps to Γ in • mode
19    end if
20    return σ
21  end def
```

Lst. 1. Instrumentation operations for direct and priority tracer modes

A Choreographed Outline Instrumentation Algorithm for Asynchronous Components

and ensures that every event message is eventually delivered. We implement hop-by-hop routing using the operations Route and Forwd (see app. A). Route creates a new *wrapper* message, $r$, with type rtd and *embeds* the message to be routed. Tracers then process routed messages by *(i)* either extracting the embedded message via the field $r$.emb, e.g. line 68 in ForwdDtc, or, *(ii)* forwarding it to the next tracer using Forwd, e.g. line 70 in ForwdDtc.

*Temporal coherence of events.* Hop-by-hop routing does not guarantee that tracers receive events in an order that reflects the correct SuS execution. This can arise when a tracer collects trace events of a SuS component *and simultaneously* receives routed events concerning this component from other tracers. Fig. 5a gives a different interleaving to the execution of fig. 4b to showcase the deleterious effect this race condition has on the runtime analysis when events are reordered for $T_Q$. In step ⑫, $T_Q$ takes the place of $T_P$ and continues tracing process $Q$, collecting the event ext in step ⑮; this happens *before* $T_Q$ receives the routed event rcv concerning $Q$ in step ⑰. When $T_Q$ analyses trace events from its message queue in the order it receives them, as in step ⑱, it violates the logical event ordering established in fig. 2b

```
 1  def Loop∘(σ,pₐ)
 2    forever do
         # Process routed messages AND direct trace events
 3      m ← next message from queue K
 4      if m.type = evt then
 5        σ ← HandleEvent∘(σ,m,pₐ)
 6      else if m.type = dtc then
           # dtc command received from issuer tracer. Route
           # dtc back to issuer through the tracer choreograpy
 7        σ ← RouteDtc∘(σ,m,pₐ)
 8      else if m.type = rtd then
 9        σ ← ForwdRtd∘(σ,m,pₐ)
10      end if
11    end forever
12  end def

13  def HandleEvent∘(σ,e,pₐ)
14    if e.act = frk then
15      σ ← HandleFork∘(σ,e,pₐ)
16    else if e.act = ext then
17      σ ← HandleExit∘(σ,e,pₐ)
18    else if e.act ∈ {snd,rcv} then
19      HandleComm∘(σ,e,pₐ)
20    end if
21    return σ
22  end def

23  def HandleFork∘(σ,e,pₐ)
24    if (pₜ ← σ.Π(e.src)) ≠ ⊥ then
25      Route(e,pₜ)
         # Route for e.tgt goes via the same tracer pₜ of e.src
26      σ.Π ← σ.Π ∪ {⟨e.tgt,pₜ⟩}
27    else
28      Delegate e to analyser pₐ OR analyse e internally
29      σ ← Instrument∘(σ,e,self())
30    end if
31    return σ
32  end def

33  def HandleExit∘(σ,e,pₐ)
34    if (pₜ ← σ.Π(e.src)) ≠ ⊥ then
35      Route(e,pₜ)
36    else
37      Delegate e to analyser pₐ OR analyse e internally
38      σ.Γ ← σ.Γ \ {⟨e.src,∘⟩}  # Remove terminated e.src
39      TryGC(σ,pₐ)
40    end if
41    return σ
42  end def

43  def HandleComm∘(σ,e,pₐ)
44    if (pₜ ← σ.Π(e.src)) ≠ ⊥ then
45      Route(e,pₜ)
46    else
47      Delegate e to analyser pₐ OR analyse e internally
48    end if
49  end def

50  def RouteDtc∘(σ,c,pₐ)
51    if (pₜ ← σ.Π(c.tgt)) ≠ ⊥ then
52      Route(c,pₜ)
53      σ.Π ← σ.Π \ {⟨c.tgt,pₜ⟩}  # Clear route for c.tgt
54      TryGC(σ,pₐ)
55    end if
56    return σ
57  end def

58  def ForwdRtd∘(σ,r,pₐ)
59    m ← r.emb
60    if m.type = dtc then
61      σ ← ForwdDtc∘(σ,r,pₐ)
62    else if m.type = evt then
63      σ ← ForwdEvt(σ,r)
64    end if
65    return σ
66  end def

67  def ForwdDtc∘(σ,r,pₐ)
68    c ← r.emb
69    if (pₜ ← σ.Π(c.tgt)) ≠ ⊥ then
70      Forwd(r,pₜ)
71      σ.Π ← σ.Π \ {⟨c.tgt,pₜ⟩}  # Clear route for c.tgt
72      TryGC(σ,pₐ)
73    end if
74    return σ
75  end def

Expect:  σ.Π(r.emb.src) ≠ ⊥
76  def ForwdEvt(σ,r)
77    e ← r.emb
78    if (pₜ ← σ.Π(e.src)) ≠ ⊥ then
79      Forwd(r,pₐ)
         # Route for e.tgt goes via the same tracer pₜ of e.src
80      if e.act = frk then
81        σ.Π ← σ.Π ∪ {⟨e.tgt,pₜ⟩}
82      end if
83    end if
84    return σ
85  end def
```

Lst. 2. Tracer loop that handles direct (∘) trace events, message routing and forwarding for our running example. Naïvely handling ext and rcv in sequence *erroneously* suggests that $Q$ receives messages after it terminates.

Tracers circumvent this issue by *prioritising* the processing of routed event messages. This captures the invariant that routed events temporally precede all other events that are to be analysed by the tracer. A tracer operates on two levels, *priority* mode and *direct* mode, denoted by ● and ○ in our algorithm. Fig. 5b shows that when in priority mode, $T_Q$ dequeues and handles the routed events rcv and frk (labelled with ●) first: rcv is analysed in step ㉓, whereas frk results in the instrumentation of tracer $T_R$ in step ㉕. Events that should not be analysed by the tracer are forwarded as outlined in fig. 4b. Note that $T_Q$ can still receive events from process $Q$ while this is underway, but these are only considered once $T_Q$ transitions to direct mode. Newly-instrumented tracers default to *priority* mode to process routed events first.

LOOP● in lst. 3 shows the logic that prioritises the processing of routed events dequeued on line 3 and handled on line 6. The operations HANDLEFORK, HANDLEEXIT and HANDLECOMM in LOOP○ and LOOP● in lsts. 2 and 3, handle trace events differently. In direct mode, a tracer can *(i)* analyse trace events, *(ii)* forward the events that have been routed its way to neighbouring tracers, or, *(iii)* start routing events that it directly collects when these need to be handled by other tracers. By contrast, tracers in priority mode only handle routed trace events according to *(i)* and *(ii)*, e.g. the branching statement on lines 24 to 31 in lst. 3, and *no* routing is performed.

*Transitioning to direct mode.* Let us refer to a *router tracer* as one that collects trace events for a system component that is meant to be traced by a *different* tracer. In fig. 4a, $T_P$ becomes the router tracer of $Q$ when it receives the events rcv and frk of $Q$, steps ⑦, ⑨, since $Q$ is meant to be traced by $T_Q$. These events cannot be handled by $T_P$, but must be routed to $T_Q$ once this is instrumented by $T_P$. Tracer may become routers depending on the interleaved executions that arise.

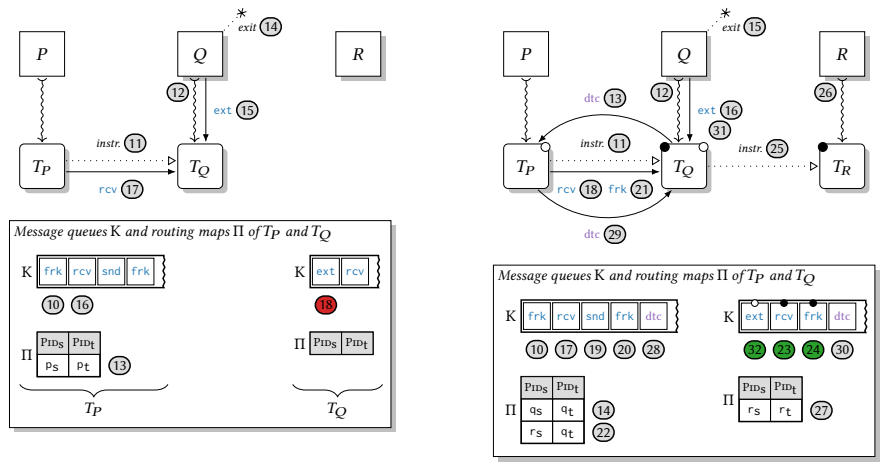

(a) $T_Q$ observes event ext before $T_P$ routes rcv  (b) $T_Q$ processes priority events routed by $T_P$ first

Fig. 5. Trace event order preservation using priority (●) and direct (○) tracer modes (analysers omitted)

A Choreographed Outline Instrumentation Algorithm for Asynchronous Components

```
 1  def Loop•(σ,pₐ)                                34  def HandleExit•(σ,r,pₐ)
 2    forever do                                   35    e ← r.emb
 3      r ← next rtd message from queue K          36    if (pₜ ← σ.Π(e.src)) ≠ ⊥ then
 4      m ← r.emb                                  37      Forwd(r,pₜ)
 5      if m.type = evt then                       38    else
 6        σ ← HandleEvent•(σ,r,pₐ)                 39      Delegate e to analyser pₐ OR analyse e internally
 7      else if m.type = dtc then                  40      σ.Γ ← σ.Γ\{⟨e.src,•⟩}   # Remove terminated e.src
         # dtc command routed back from router tracer 41    TryGC(σ,pₐ)
 8        σ ← HandleDtc(σ,r,pₐ)                    42    end if
 9      end if                                     43    return σ
10    end forever                                  44  end def
11  end def
                                                   45  def HandleComm•(σ,r,pₐ)
12  def HandleEvent•(σ,r,pₐ)                       46    e ← r.emb
13    e ← r.emb                                    47    if (pₜ ← σ.Π(e.src)) ≠ ⊥ then
14    if e.act = frk then                          48      Forwd(r,pₜ)
15      σ ← HandleFork•(σ,r,pₐ)                    49    else
16    else if e.act = ext then                     50      Delegate e to analyser pₐ OR analyse e internally
17      σ ← HandleExit•(σ,r,pₐ)                    51    end if
18    else if e.act ∈ {snd,rcv} then               52  end def
19      HandleComm•(σ,r,pₐ)                        Expect:  r.emb.iss = self() ∨ σ.Π(r.emb.tgt) ≠ ⊥
20    end if                                       53  def HandleDtc(σ,r,pₐ)
21  end def                                        54    c ← r.emb
                                                   55    if (pₜ ← σ.Π(c.tgt)) ≠ ⊥ then
22  def HandleFork•(σ,r,pₐ)                        56      Forwd(r,pₜ)
23    e ← r.emb                                    57    else
24    if (pₜ ← σ.Π(e.src)) ≠ ⊥ then                58      σ.Γ ← σ.Γ\{⟨c.tgt,•⟩}
25      Forwd(r,pₜ)                                59      σ.Γ ← σ.Γ∪{⟨c.tgt,○⟩}
26      σ.Π ← σ.Π∪{⟨e.tgt,pₜ⟩}                     60      γ = {⟨pₛ,d⟩ | ⟨pₛ,d⟩ ∈ σ.Γ, d = •}
27    else                                         61      if γ = ∅ then   # Check all processes in Γ are detached
28      Delegate e to analyser pₐ OR analyse e internally  62  Loop○(σ,pₐ)  # Switch tracer to ○ mode
29      p'ₜ ← r.rtr                                63      end if
30      σ ← Instrument•(σ,e,p'ₜ)                   64    end if
31    end if                                       65    return σ
32    return σ                                     66  end def
33  end def
```

Lst. 3. Tracer loop that handles priority (•) trace events and message forwarding

A tracer in priority mode coordinates with its router tracer to determine when all of the events for a system process it traces have been routed to it. Each tracer keeps a record of the processes it traces in the *traced-processes map*, $\Gamma : \text{PID}_s \rightharpoonup \{\circ, \bullet\}$. Entries to $\Gamma$ are added when the tracer starts collecting events for a process (lines 7 and 18 in lst. 1) and removed when processes terminate (lines 38 and 40 in lst. 2 and lst. 3). The coordination procedure with the router is effected by the tracer in priority mode, once for *every* process in $\Gamma$, before it can safely transition to direct mode and start operating on the events it collects directly. The tracer issues a special detach command message, $c$, with type dtc, to notify the router tracer that it is now responsible for tracing a particular system process. The detach command contains the PIDs of the issuer tracer and system process, read via the respective fields $c$.iss and $c$.tgt. Tracers mark a process as detached by updating its mapping $c.\text{tgt} \mapsto \bullet$ in $\Gamma$ to $c.\text{tgt} \mapsto \circ$.

Fig. 5b shows $T_Q$ in priority mode sending command $\langle \text{dtc}, q_t, q_s \rangle$ for $Q$, step ⑬, after it starts tracing this process in step ⑫. This transaction is implemented by Detach on line 17 in lst. 1 (see app. A). The dtc command issued by $T_Q$ is deposited in the message queue of (router tracer) $T_P$ after the events rcv and frk. $T_P$ processes the contents of its message queue sequentially in steps ⑩, ⑰, ⑲, ⑳ and ㉘, and forwards rcv and frk, steps ⑱ and ㉑. It also routes the dtc command *back* to the issuer tracer $T_Q$, step ㉙, where once handled, marks $Q$ as detached from $T_P$. This update on $\Gamma$ of $T_Q$ is performed by HandleDtc in lst. 3 on lines 58 and 59.

A tracer transitions to direct mode once *all* the processes in $\Gamma$ become detached; see lines 60 and 61 in lst. 3. While in priority mode, $T_Q$ in fig. 5b handles the events forwarded by $T_P$ in the correct order, as per fig. 2b (steps ㉓ and ㉔).

This is followed by handling dtc in step ㉚. The transition from priority to direct mode for $T_Q$ in fig. 5b takes place in step ㉛. Finally, the event ext is handled in the correct order in step ㉜ (as opposed to step ⑱ in fig. 5a).

A detach command ⟨dtc,$p_t$,$p_s$⟩ that is routed to a tracer $p_t$ by the router, may be directed via multiple *intermediate tracers* before it reaches $p_t$. Every tracer *en route* to $p_t$ purges the mapping for $p_s$ from its routing map Π once it forwards dtc to the next tracer. This cleanup logic is performed by RouteDtc and ForwdDtc in lst. 2. Fig. 5b does not illustrate this flow. We however note that $T_P$ would remove from Π the mapping $q_s \mapsto q_t$, calling RouteDtc to route back the detach command ⟨dtc,$q_t$,$q_s$⟩ it receives from $T_Q$. Eventually, $T_P$ also removes $r_s \mapsto q_t$ for $R$ once it handles ⟨dtc,$r_t$,$r_s$⟩ from $T_R$. When $T_Q$ receives the *rerouted* detach command ⟨rtd,$p_t$,⟨dtc,$r_t$,$r_s$⟩⟩ from $T_P$, it removes $r_s \mapsto r_t$ from Π and forwards it, in turn, to $T_R$.

*Selective instrumentation.* To monitor processes as one component—rather than having a dedicated monitor for each as in our example—the algorithm uses the *instrumentation map*, $\Phi : \text{Sig} \rightharpoonup \text{Mon}$, from code signatures, $g$, to analysis code, $v$. $\Phi$ is consulted to selectively instrument processes. The signature $g$, carried as part of the fork trace event $e$, can be retrieved using the field $e$.sig; see tbl. 1. The instrumentation operations in lst. 1 apply $\Phi$ to $e$.sig on lines 3 and 13. When $\Phi(e.\text{sig}) = \bot$, no instrumentation is effected, and the tracer becomes automatically shared by the new process $e$.tgt, as per assumptions $A_8$ and $A_9$.

*Garbage collection.* Our outline instrumentation can shrink the tracer choreography by discarding unneeded tracers. A tracer is designed to terminate when both its routing Π *and* traced-processes Γ maps become empty. It purges process references from Γ when handling exit trace events via HandleExit$_\circ$ and HandleExit$_\bullet$ (lsts. 2 and 3). When Γ = ∅, and a tracer has no processes to analyse, it could still be required to forward events to neighbouring tracers, *i.e.,* Π ≠ ∅. Therefore, the garbage collection check, TryGC, is performed each time mappings from Π or Γ are removed; see lines 39, 54 and 72 in lst. 2, and line 41 in lst. 3.

## 3 CORRECTNESS VALIDATION

Our choreographed instrumentation from sec. 2 is assessed in two stages. First, we confirm its *implementability* by instantiating the core logic of lsts. 1–3 to a multipurpose language that is tailored to the demands of distributed applications. Our development follows a test-driven approach [13, 85], to ensure that the tracer logic is implemented correctly. Second, we verify the correctness of our implementation by augmenting the logic given in lsts. 1–3 with runtime checks that guarantee a number of invariants [8, 75] w.r.t. message routing between tracers.

*Implementability.* We map our algorithm of sec. 2.4 to the actor model of computation [4, 49]. *Actors*—the basic unit of decomposition in this model—are abstractions of concurrent entities that do not share mutable memory with other actors (assumption $A_1$). Instead, actors interact through *asynchronous messaging*, and alter their internal state based on messages they consume (assumption $A_5$). Each actor is equipped with an incoming message buffer called the *mailbox*, from where messages deposited by other actors may be *selectively* read (assumption $A_4$). Besides sending and receiving messages, actors can fork other actors (assumption $A_2$). Actors are uniquely identifiable via their PID that they use to directly address one another (assumption $A_3$).

The actor model is concretised by a number of languages and frameworks, including Erlang, Akka for the Java Virtual Machine (JVM), Thespian [69] for Python [59], and Pony [84]. We adopt Erlang for our implementation, since the Erlang Virtual Machine (EVM) is *specifically engineered* for high-concurrency settings Armstrong [6], Cesarini and Thompson [24]: it implements actors as lightweight processes, permitting us to scale our experiment set-ups

A Choreographed Outline Instrumentation Algorithm for Asynchronous Components

to loads that go *beyond* the state of the art. The EVM also features *per-process* garbage collection that—unlike JVM implementations—does not subject the entire virtual machine to non-deterministic pauses [52, 67]; this minimises the interference with running benchmarks and is fundamental to stabilising the *variance* in our measurements. All these aspect increases the confidence in our empirical measurements and the general conclusion drawn from them.

Our implementation of choreographed outline monitoring maps the tracer processes of sec. 2 to Erlang actors. Every tracer collects events from components of the SuS by leveraging the *native tracing infrastructure* of the EVM [24], which complies with assumptions $A_7$–$A_9$. It deposits trace event messages inside the tracer mailboxes, which in our implementation, coincide with the message queues K of sec. 2.2. Our outline set-up covers both the externalised (EA) and internalised (IA) analysis variants of fig. 1[1].

*Invariant and unit testing.* One salient aspect our algorithm addresses is that of reporting SuS trace events to the analysis component in a *reliable* manner; this is captured by requirement $R_3$. The invariants listed below ensure the correct handling of events by tracers. Together with the core logic of lsts. 1–3, these enable us to reason about properties the tracer choreography should observe. For instance, our algorithm guarantees that 'every trace event that is routed between tracers eventually reaches the intended tracer', that 'the choreography grows dynamically', and that 'unneeded tracers are always garbage collected'. We implement these invariant checks in the form of assertions as part of our implementation.

**Choreography invariants**

$I_1$ The root tracer has *no* router tracers.

$I_2$ A tracer $p_t$ has exactly *one* router tracer $p'_t$; this does not apply for the root tracer, invariant $I_1$. The router tracer of $p_t$ is either its parent *or* the tracer that forked the ancestors of $p_t$.

$I_3$ A tracer never terminates unless its routing map, $\Pi$, *and* traced-processes map, $\Gamma$, are empty.

$I_4$ A tracer that analyses a frk event adds a process to its traced-process map $\Gamma$. Depends on invariant $I_{25}$.

$I_5$ A tracer that analyses an ext event removes a process from its traced-processes map $\Gamma$. Depends on invariant $I_{26}$.

$I_6$ A tracer that analyses a frk event adds a route to its routing map $\Pi$. Depends on invariant $I_{23}$.

$I_7$ A tracer that routes a frk event adds a route to its routing map $\Pi$. Depends on invariant $I_{23}$.

$I_8$ A tracer that forwards a frk event adds a route to its routing map $\Pi$. Depends on invariant $I_{23}$.

$I_9$ A tracer that routes a dtc command removes a route from its routing map $\Pi$. Depends on invariant $I_{24}$.

$I_{10}$ A tracer that forwards a dtc command removes a route from its routing map $\Pi$. Depends on invariant $I_{24}$.

**Message routing invariants**

$I_{11}$ A tracer in ● mode prioritises routed messages until it switches to ○ mode.

$I_{12}$ A tracer handles a dtc command only in ● mode.

$I_{13}$ A tracer in ● mode either analyses a routed trace event *or* forwards it. Depends on invariants $I_{21}$ and $I_{22}$. A tracer can only route direct events, *i.e.,* when $m$.type = rtd, in ○ mode. Routing in ● mode means that the tracer dequeued a direct event, violating invariant $I_{11}$.

$I_{14}$ A tracer in ● mode either handles a routed dtc command *or* forwards it. Depends on invariants $I_{10}$ and $I_{22}$. A tracer can only route direct commands, *i.e.,* when $m$.type = rtd, in ○ mode. Routing in ● mode means that the tracer dequeued a direct command, violating invariant $I_{11}$.

$I_{15}$ A tracer in ○ mode either analyses a direct trace event *or* routes it. Depends on invariants $I_{21}$ and $I_{22}$.

---
[1]The full source code can be found on the GitHub repository: https://github.com/duncanatt/detecter.

- $I_{16}$ A tracer in ∘ mode only forwards a routed trace event, *i.e.,* when $m$.type=rtd. Depends on invariant $I_{22}$. Analysing a routed trace event in ∘ mode means that the tracer dequeued a priority event, violating invariant $I_{11}$.
- $I_{17}$ A tracer in ∘ mode only forwards a routed dtc command, *i.e.,* when $m$.type = rtd. Depends on invariants $I_{10}$ and $I_{22}$. Handling a routed command in ∘ mode means that the tracer dequeued a priority event, violating invariants $I_{11}$ and $I_{12}$.
- $I_{18}$ A router tracer that receives a dtc command must route it. Depends on invariants $I_9$ and $I_{22}$. If routing is not possible, the command was sent by mistake.
- $I_{19}$ A tracer in • mode transitions to ∘ mode only when all of the processes in its traced-processes map $\Gamma$ are marked as ∘ or $\Gamma$ is empty. Depends on invariant $I_{12}$.
- $I_{20}$ The total amount of dtc commands a tracer issues is equal to the sum of the number of processes in its traced-process map $\Gamma$ *and* the number of terminated processes for the tracer. Depends on invariants $I_4$ and $I_5$.
- $I_{21}$ A tracer has a corresponding analyser.
- $I_{22}$ A tracer never routes *or* forwards a message unless a route exists in its routing map $\Pi$. Depends on invariants $I_6$–$I_8$.
- $I_{23}$ A tracer never adds a route that already exists in its routing map $\Pi$.
- $I_{24}$ A tracer never removes a non-existing route from its routing map $\Pi$.
- $I_{25}$ A tracer never adds a process that already exists in its traced-processes map $\Gamma$.
- $I_{26}$ A tracer never removes a non-existing process from its traced-processes map $\Gamma$.

We also implement a number of unit tests that operate on these invariants. The tests focus on race conditions that arise and how these are handled by the tracer choreography while it simultaneously analyses trace events. Our tests also validate the scaling aspect of our algorithm in terms of the dynamic instrumentation of tracers and corresponding garbage collection. To this end, we built a testing harness that can be *controlled* to yield particular interleaved executions. The harness adheres to assumptions $A_7$–$A_9$, and emulates the native EVM tracing infrastructure. It enables us to inject our choreographed outline instrumentation with different *trace permutations* to test and obtain *full* branch coverage of the core logic in lsts. 1–3.Note that our tests use only valid trace permutations, following assumption $A_4$. While tracer and analyser components are not permitted to fail (assumption $A_6$), we introduce random faults to verify that *partial-failure* is possible for independently-executing tracers. We also simulate SuS components crashes to confirm that abnormal termination ext events are captured and correctly analysed by tracers.

## 4 EMPIRICAL EVALUATION

The litmus test assessing the utility of our choreographed outline instrumentation should show that it induces feasible runtime overhead. Concretely, we employ a benchmarking tool that evaluates the scalability of our implementation from sec. 3 over a master-worker SuS that is subject to a variety of loads covering most real-world scenarios.

### 4.1 Methodology

*Benchmarking set-up.* We adopt a concurrent benchmarking framework [3] written for the analysis of highly-scalable Erlang systems. The framework generates system models based on the master-worker architecture through configurable parameters. These models have been shown to be *realistic* [72] and satisfy assumptions $A_1$–$A_6$ from sec. 2. The benchmarking framework permits us to emulate different load profiles that are observed in practice, in contrast to established benchmarking tools, *e.g.* SPECjvm2008 [83], DaCapo [15], ScalaBench [79] and Savina [50], and industry-strength load testers, *e.g.* Tsung [66] or JMeter [5]. *Steady* loads reproduce executions where a system operates under stable conditions. *Pulse* loads emulate scenarios where a system undergoes gradually-increasing load peaks. *Burst* loads capture scenarios where a system is stressed due to instant load spikes. The benchmarking framework also captures four performance metrics, namely the: *(i)* mean execution duration, measured in seconds (s), *(ii)* mean scheduler utilisation, as a percentage of the total available capacity, *(iii)* mean memory consumption, measured in GB, and, *(iv)* mean response time (RT), measured in milliseconds (ms). These provide a *multi-faceted* view of runtime overhead that is conducive to a comprehensive assessment of decentralised monitoring.

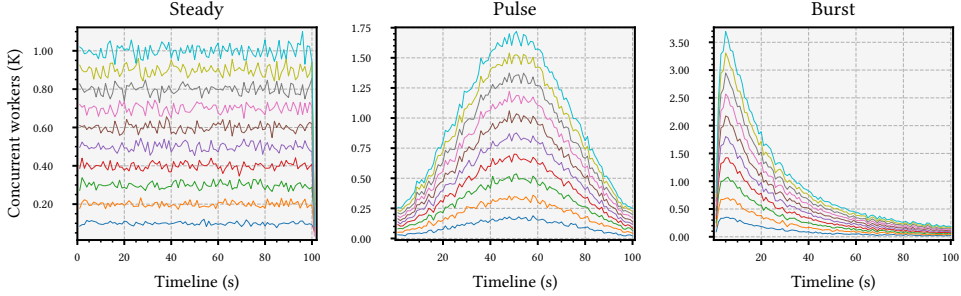

Fig. 6. Steady, Pulse and Burst load distributions with 100 k workers for the duration of 100 s

*Experiment set-up.* We carry out two case studies, instrumenting the master-worker SuS with either inline or outline monitors to compare them with the uninstrumented SuS as a baseline. The first case study configures the SuS with 10 k for *moderate* loads, whereas the second case study uses 100 k workers for *high* loads. In each case, the master issues 100 requests to each worker; these requests are distributed to workers based on the Steady, Pulse and Burst load profiles (see fig. 6). The loading time is set to 100 s. For each experiment, we run 10 benchmarks using a specific monitoring set-up (*e.g.* outline instrumentation), incrementally applying load in steps. We perform 10 repetitions of the same experiment and aggregate the results for the four performance metrics recorded by the benchmarking framework. Our experiments are conducted on an Intel Core i7 M620 64-bit machine with 8GB of memory, running Ubuntu 18.04 LTS and Erlang/OTP 22.2.1.

*Inline monitoring.* We are unaware of inline monitoring tools that target programs written for the Erlang platform. As a result, our study required us to develop a bespoke inline monitoring tool that models Parametric Trace Slicing [10, 26]. It instruments monitoring instructions into the target program via code injection by manipulating its abstract syntax tree. The modified syntax tree is afterwards compiled, and the weaved instructions effect the runtime analysis in a *synchronous* manner. Further detail on the tool may be found in app. B.

### 4.2 Precautions

Recall that RV monitoring is made up of two parts: instrumentation and analysis, *e.g.* fig. 1. We objectively compare inline and outline instrumentation by fixing the runtime overhead introduced by the analysis. To this end, our synthesis is designed to generate *identical* analysis components for both the inline and outline implementations. The analysis components process trace event streams and reach verdicts in connection with correctness properties [11, 43]. Our synthesis generates *automata-based* analysers from syntactic descriptions of properties, following the approach of established RV tools such as Aceto et al. [1], Cassar et al. [22], Havelund and Rosu [48], Meredith et al. [60], Mostafa and Bonakdarpour [62], Reger et al. [70]. Unfortunately, this is not enough to guarantee a constant analysis overhead across inline and outline instrumentations. Since inline and outline monitors behave differently (e.g. inline monitors can query the internal state of the SuS, whereas outline monitors must replicate this externally), this disparity introduces runtime biases that obstruct the collection and isolation of instrumentation overheads, complicating our interpretation of results. To minimise these effects, our case studies employ properties that are *parametric* w.r.t. SuS components, where the synthesised analysers are able to reach verdicts *without* mutually interacting e.g. Attard and Francalanza [7], Chen and Rosu [26], Jin et al. [51], Neykova and Yoshida [64], Reger et al. [70].

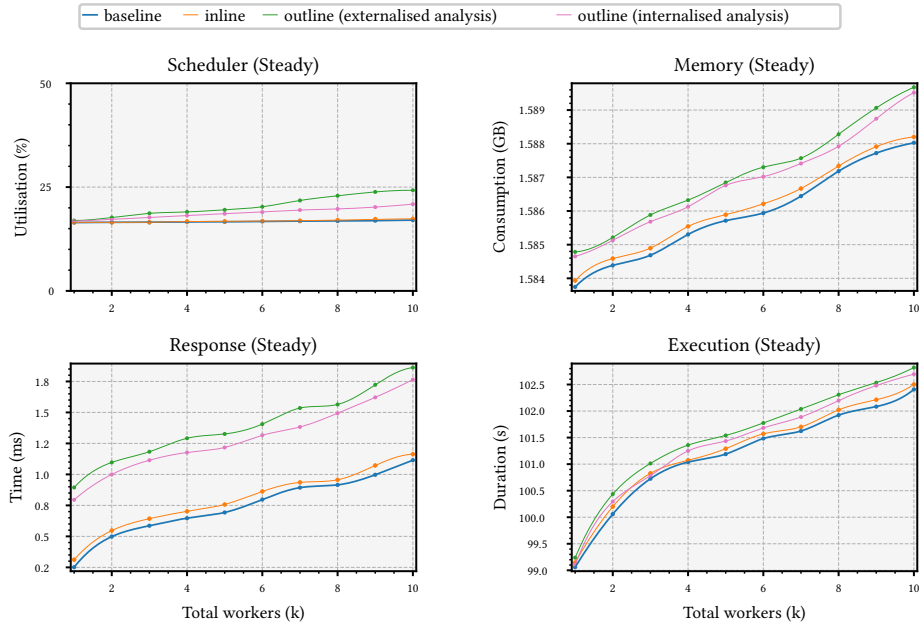

Fig. 7. Mean runtime overhead for benchmark configured with moderate load (10k workers)

### 4.3 Results and Discussion

We perform two sets of experiments, for *moderate* and *high* loads. Our results are reported in tbls. 2 and 3 and figs. 7–10. Charts plot each performance metric (*y*-axis) against the number of worker processes (*x*-axis) for the inline and the two outline instrumentation variants of fig. 1, EA and IA. The *unmonitored* system is inserted as a baseline reference.

*Moderate loads.* This first batch of results replicates loads that are slightly higher than those used by the state of the art in decentralised (and distributed) RV e.g. Attard and Francalanza [7], Berkovich et al. [14], Cassar and Francalanza [21], Colombo and Falcone [29], El-Hokayem and Falcone [36], Francalanza and Seychell [45], Mostafa and Bonakdarpour [62], Neykova and Yoshida [64, 65], Scheffel and Schmitz [73]. The set-up with 10k workers and 100 work requests generates $\approx 10\text{k} \times 100 \times$ (work requests and respones) = 2M message exchanges, producing 2M× (snd and rcv trace events) = 4M analysable events. Tbl. 2 reports the percentage overhead at 10k workers. It shows that inline and the two variants of outline monitoring, EA and IA, induce negligible execution slowdown for *all* three load profiles. For instance, the maximum slowdown for EA is 1.09 %; the memory consumption behaves similarly. At the Steady load illustrated in fig. 7, the memory consumption and RT grow *linearly* in the number of worker processes. For the three load profiles in tbl. 2, inline monitoring induces minimal scheduler overhead; this is markedly higher for EA and IA, and is mostly caused by the dynamic reconfiguration of the monitoring choreography. Tbl. 2 indicates that the RT is *sensitive* to the type of load applied, where it increases under the Steady, Pulse and Burst load profiles respectively. Bursts, in fact, induce a *sharp growth* in the RT for outline monitoring at 9k ∼ 10k workers, as fig. 8 shows.

We stress that—notwithstanding the clear percentage-wise discrepancies in the scheduler utilisation and RT between inline and outline monitoring in tbl. 2—this overhead is *comparable value-wise* for moderate loads that are typically used in other bodies of work. Concretely, fig. 8 shows that the worst discrepancy for the RT at a Burst load of 10k is a mere 11ms. Merging the tracing and analysis (IA) does yield improvements, but the effect is minor at this load size. Figs. 7 and 8 also show that the memory consumption, RT and execution duration plots exhibit *analogous growth* for inline and outline monitoring under the Steady and Pulse load profiles, differing slightly in the *y*-intercept value. This is a good indicator that for moderate loads *both* forms of instrumentation scale well for these profiles.

We remark that the RT Burst plots for EA and IA in fig. 8 prevent us from extrapolating our findings. In fact, these attest to the utility of benchmarks using different forms of load that, typically, do not feature in other studies; see Aceto et al. [3] for details. Coupled with a multi-faceted view of overhead, load profiles may reveal *nuanced behaviour* that could be overlooked, were one to consider a *single* performance metric (*e.g.* execution slowdown). For our moderate loads, the RT Burst plots raise the question of whether the observed trend for EA and IA remains consistent when the load goes beyond 10k workers.

*High loads.* We assesses inline and outline monitoring under high concurrency. The maximum number of workers is set to 100k to produce 20M messages and 40M trace events. Fig. 6 shows that the benchmark with a Steady load of 100k generates an average of 1k workers/s; this load spikes to just over 3.5k workers for Burst.

Tbl. 3 confirms that inline monitoring does induce lower overhead. However, dissecting these results uncovers important subtleties. For instance the difference in memory consumption between inline and EA is 13.3 % under a

|  | Steady | | | Pulse | | | Burst | | |
| --- | --- | --- | --- | --- | --- | --- | --- | --- | --- |
|  | Inline | EA | IA | Inline | EA | IA | Inline | EA | IA |
| **Scheduler utilisation** | 1.68 | 42.13 | 22.60 | 1.54 | 35.17 | 18.12 | 2.08 | 38.92 | 26.03 |
| **Memory consumption** | 0.01 | 0.10 | 0.09 | 0.01 | 0.08 | 0.04 | 0.03 | 0.10 | 0.06 |
| **RT** | 4.37 | 67.05 | 58.36 | 7.72 | 82.85 | 60.79 | 20.17 | 859.91 | 666.46 |
| **Execution duration** | 0.09 | 0.40 | 0.28 | 0.10 | 0.32 | 0.22 | 0.12 | 1.09 | 0.77 |

Tbl. 2. Percentage runtime overhead w.r.t. to baseline at the *maximum* load of 10k workers

|  | Steady | | | Pulse | | | Burst | | |
| --- | --- | --- | --- | --- | --- | --- | --- | --- | --- |
|  | Inline | EA | IA | Inline | EA | IA | Inline | EA | IA |
| **Scheduler utilisation** | 1.8 | 86.8 | 58.1 | 2.9 | 85.5 | 55.6 | 3.1 | 84.4 | 57.9 |
| **Memory consumption** | 1.9 | 15.2 | 8.7 | 2.9 | 18.1 | 11.8 | 3.1 | 24.6 | 15.4 |
| **RT** | 68.9 | 326.9 | 267.9 | 72.7 | 257.8 | 238.7 | 28.4 | 120.6 | 114.3 |
| **Execution duration** | 23.5 | 108.6 | 93.8 | 24.5 | 101.8 | 93.5 | 15.7 | 82.0 | 77.5 |

Tbl. 3. Percentage runtime overhead w.r.t. to baseline at the *maximum* load of 100k workers

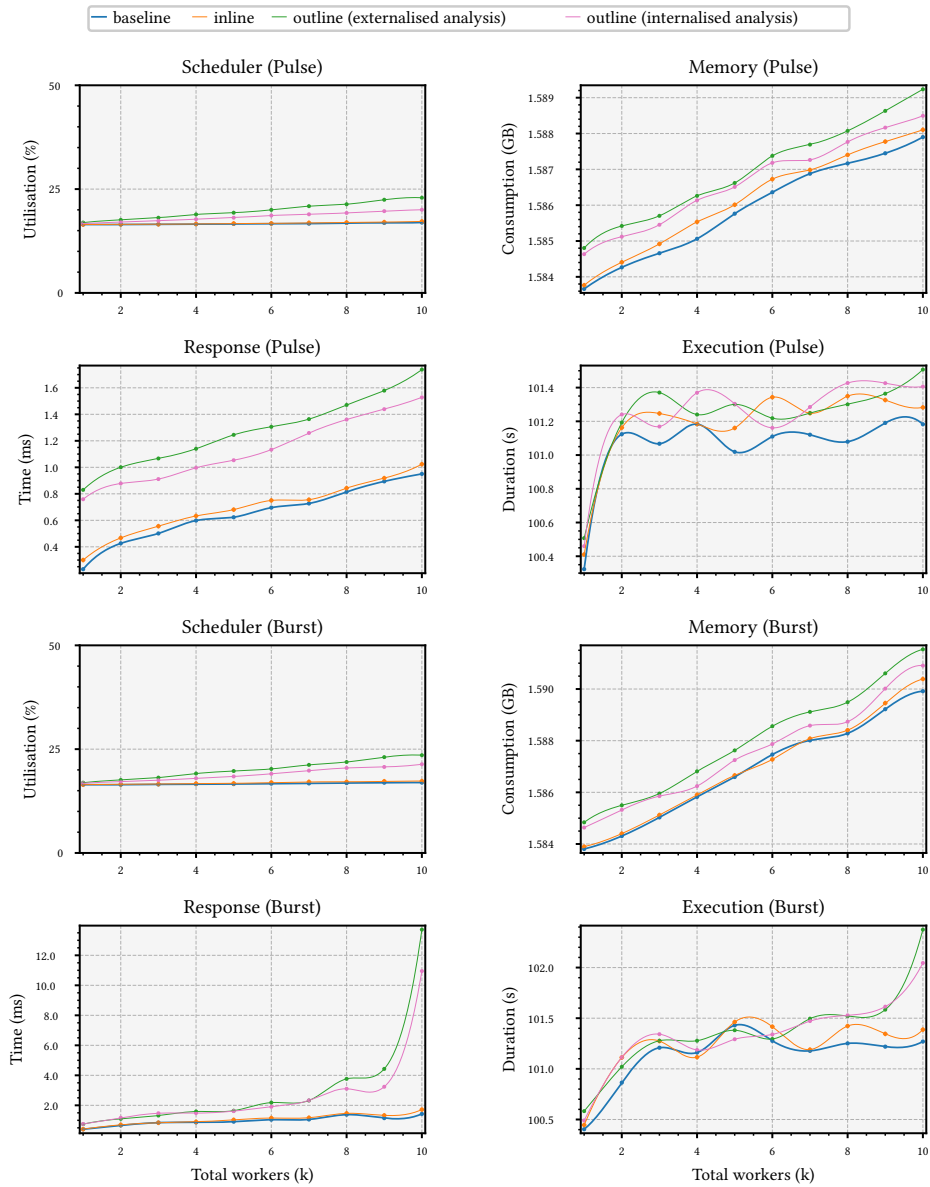

Fig. 8. Mean runtime overhead for benchmark configured with moderate load (10k workers, cont.)

Steady load; localising the analysis (IA) reduces this gap to 6.8 %. This overhead is arguably tolerable for a number of applications, and *debunks* the general notion that outline monitoring necessarily leads to infeasible overhead. Fig. 9 also shows that under a Steady load, the memory consumption, RT, and execution duration overhead induced by outline monitoring is similar to those of inline monitoring up to the respectable load of ≈ 40k workers, *i.e.,* 8M messages and 16M events. We point out the benefit that IA yields when compared to EA, discernible in the form of *consistently* lower overhead for *all* four performance metrics, irrespective of the load profile.

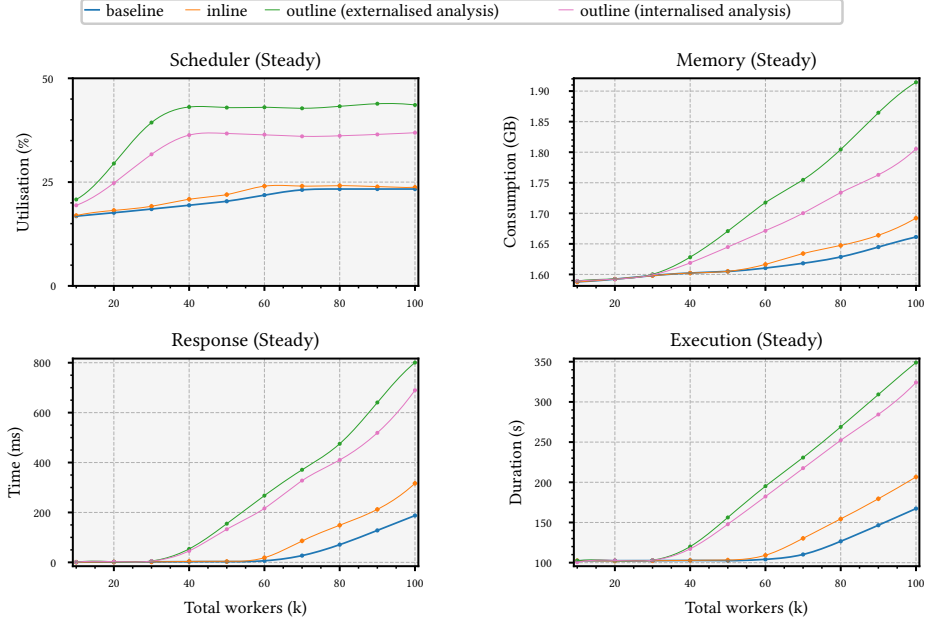

Fig. 9. Mean runtime overhead for benchmark configured with high load (100k workers)

The memory consumption, RT, and execution duration plots in figs. 9 and 10 exhibit a *linear* growth beyond specific $x$-axis thresholds. This contrasts with the plots in figs. 7 and 8 for 10k workers, where varying trends may be observed. Specifically, the execution duration under the Steady load profile appears to grow (negative) quadratically, but follows a cubic shape in the case of Pulse and Burst loads; a similar effect is obtained in the RT under the Burst load profile.

*Estimated overhead on workers.* We refine the results in tbl. 3 and estimate the overhead at each worker for *outline monitoring*. Qualifying this overhead is crucial when the processing capability of components is spread on nodes with differing resource requirements. For example, in a set-up where workers carry out computationally-intensive tasks (*e.g.* render farms [87]), gauging the RT between components may be more relevant than measuring the scheduler utilisation; in an IoT scenario with low-powered devices (*e.g.* sensor data acquisition [27]), calculating the resource usage is useful to predict battery consumption. Estimating the overhead per worker also gives a *truer depiction* of its distribution, in contrast to a global view. In a master-worker architecture, where the master is susceptible to bottlenecks, a large portion of overhead is incurred by the master [72], yet this is not clearly perceivable in the presented plots. Our chosen benchmarking tool does not enable us to measure the overhead at each worker. We approximate this by subtracting the total overhead induced by outline monitors from the overhead at the master, dividing the difference by the number of workers, *i.e.,* 100k. The result obtained for the four performance metrics is negligible—at < 0.01%—for

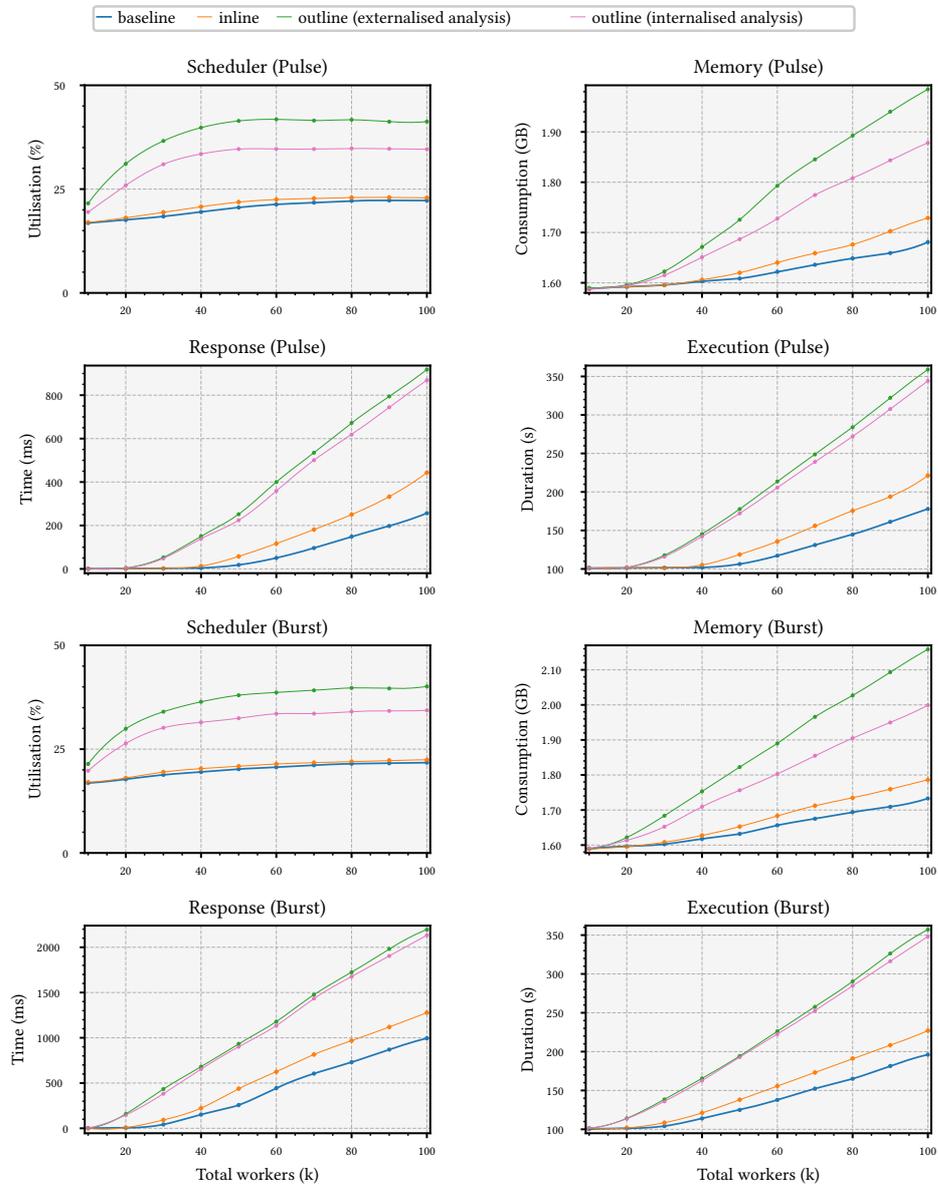

Fig. 10. Mean runtime overhead for benchmark configured with high load (100k workers, cont.)

Steady, Burst and Pulse loads, demonstrating that inline and outline monitoring induce much the same overhead (see tbl. 4 in app. C).

## 5 CONCLUSION AND RELATED WORK

This paper proposes a new instrumentation approach—*choreographed outline instrumentation*—targeted towards the monitoring of decentralised, open systems. Sec. 2 details a concrete algorithm that describes how the instrumentation of a concurrent SuS is attainable in a decentralised, scalable fashion by relying exclusively on events exhibited by the running system; contribution (i). Our approach sets itself apart from the state of the art in three core aspects: *(i)* it *asynchronously* instruments the SuS with monitors *without* modifying it, *(ii)* *dynamically reorganises* the monitoring choreography while the runtime analysis is in progress, and, *(iii)* does *not* assume a fixed number of SuS components, but *scales* as required. The exposition in sec. 2 identifies the intricacies that our algorithm addresses in order to guarantee that trace events of the SuS are reported *and* analysed correctly, lsts. 1–3. We express our algorithm in terms of general software engineering concepts (*e.g.* encapsulated component states, separation of the routing and analysis concerns) to facilitate its adoption to a variety of settings and technologies. The algorithm we present is evaluated in two respects. Sec. 3 confirms the *implementability* of choreographed outline instrumentation; contribution (ii). It describes how our general algorithm of lsts. 1–3 can be naturally mapped to a tool implementation in a mainstream concurrent language. We augment this with an account of the principled approach employed to ensure the *correct* translation of our algorithm to code. Sec. 4 assesses the *feasibility* of choreographed outline instrumentation in terms of runtime overhead; contribution (iii). We subject our implementation of sec. 3 to rigorous empirical analyses to show that the overhead it induces is *comparable* to that of inline instrumentation for many practical cases.

*Impact.* We debunk the commonly-held view that outline monitoring necessarily induces infeasible overhead. Our study shows that, in the case of decentralised systems, outline monitoring can be a viable alternative, even when conducted in decentralised fashion. To conclude its feasibility, our study considers other facets of overhead apart from the execution slowdown, and shows that outline monitoring induces comparable overhead to inlining in the case of master-worker systems. We conjecture that, in other architectures where the distribution of load is more even (*e.g.* peer-to-peer), this discrepancy is even less. Moreover, our experiments subject the monitoring set-up to different load profiles that are typically observed in practice, fig. 6, increasing our confidence in the applicability of our results; this is generally not done by other studies, *e.g.* Attard and Francalanza [7], Berkovich et al. [14], Bodden et al. [16], Cassar and Francalanza [21], Cassar et al. [23], Chen and Rosu [25, 26], Colombo et al. [30, 31, 32], El-Hokayem and Falcone [36, 37], Francalanza and Seychell [45], Lange and Yoshida [55], Meredith et al. [60], Mostafa and Bonakdarpour [62], Neykova and Yoshida [64, 65], Reger et al. [70]. Although the empirical measurements presented in sec. 4 are necessarily dependent on the chosen implementation framework, our conclusions should be transferrable to other set-ups that adhere to assumptions $A_1$–$A_9$.

Our solution adopts a principle similar to the black-box-style of monitoring employed by Application Performance Management (APM) tools that are geared towards maintaining large-scale decentralised software. APM tools operate externally to the SuS, similar to our tool. They are used extensively to identify and diagnose performance problems such as bottlenecks and hotspots; they presently have an edge on static analysis tools for critical path analysis [86] and unearthing performance anti-patterns [81, 82]. The proposed methods in sec. 2 are general enough to be applied, at least in part, to APM tools in order to make them more decentralised. In turn, we can also benefit from judiciously introducing advanced APM techniques such as sampling to further lower overhead. Although our algorithm is implement in Erlang [6], it is still sufficiently general to be instantiated to other language frameworks (*e.g.* Elixir [53], Akka [71] for Scala [68], Thespian [69] for Python [59]) that follow assumptions $A_1$–$A_9$. It can then, be used by RV tools that target other platforms, such as the JVM.

Hyper-logics [28] have recently emerged as a expressive formalism for expressing complex properties about decentralised systems (*e.g.* non-interference). Broadly, these logics can specify conditions across distinct traces, where quantifications range over potentially infinite trace domains. One branch in this line of study is the verification of such properties at runtime [17, 41]. Although we are unaware of any attempts at runtime verifying such properties using outline instrumentation, the inherent dynamicity required to analyse an unbounded number of traces would certainly make our instrumentation methods applicable. Our tool from sec. 3 already disentangles the instrumentation from the analysis, thus providing a platform for plugging new analyses that implement monitoring for hyper-properties (instead of the existing one).

*Related work.* There are other bodies of work that address decentralised monitoring besides the ones already discussed. A considerable portion of these instrument monitors via inlining. For instance, Sen et al. [78] study decentralised monitors that are attached to different threads to extract and analyse trace events internally; see fig. 1b. In their earlier work [77], the authors conduct investigate the use of decentralised monitors on distributed SuS components. While both works focus on the efficiency of communication between monitors, these do not study, nor quantify the overhead induced by runtime monitoring. Minimising overhead is also the focus of Mostafa and Bonakdarpour [62]. In their setting, the SuS consists of distributed asynchronous processes that interact via message-passing over reliable channels. Similar to ours, their monitoring algorithm does not rely on a global notion of timing (assumption $A_1$), nor does it tackle aspects of failure (assumptions $A_4$ and $A_6$). The work by Basin et al. [12] is one of the few that considers distributed system monitoring where components and network links may fail. While their algorithm does not employ a global clock, it is based on the timed asynchronous model for distributed systems [33] that assumes *highly-synchronised* physical clocks across nodes. In a different spirit, Bonakdarpour et al. [18], Fraigniaud et al. [42] address the problem of crashing monitors; this is something that we presently do not address, although our decentralised set-up enables us to *fail partially* (see sec. 3).

Other tooling efforts for decentralised monitoring such as El-Hokayem and Falcone [36], Jin et al. [51], Kim et al. [54], Sen et al. [76], weave the SuS with code instructions that extract trace events and delegate their analysis to independent processes—this mirrors our externalised event analysis variant of fig. 1b. Although these approaches may be classified under outline instrumentation [40], they *do not* treat the SuS as a black-box, which makes them prone to shortcomings of inlining. Crucially, the aforementioned works assume a *static* system arrangement, sparing them from dealing with the dynamic reconfiguration of outline tracers and reordering of tracer events.

Tools such as Attard and Francalanza [7], Neykova and Yoshida [64] target the Erlang ecosystem. [64] propose a method that statically analyses the program communication flow that is specified in terms of a multiparty protocol. Monitors attached to system processes then check that the messages received coincide with the projected local type (similar to the analysis conducted by our monitors) in the case of failure, the associated processes are restarted. The authors show that their recovery algorithm induces less communication overhead, and improves upon the static process structure recovery mechanisms offered by the Erlang/OTP platform. Similarly, [7] focus on decentralised outline monitoring in a concurrent setting. By contrast to [64], they leverage the native tracing infrastructure offered by the EVM.

We remark that the tools discussed in this section rely on bespoke evaluation platforms, making it hard to reproduce and objectively compare to ours. Their empirical measurements focus exclusively on the execution slowdown. By comparison, our measurements of sec. 4 show that this metric alone can be misleading since it does not portray the other facets of overhead, *e.g.* mean response time, that are relevant to decentralised component settings [3]; see discussion in

A Choreographed Outline Instrumentation Algorithm for Asynchronous Components

sec. 4.3. Moreover, their experiments either use a Steady load profile (*i.e.,* a Poisson process), or fail to specify the load type altogether. Our empirical study shows that different load profiles are essential to holistically assesses overhead in decentralised settings.

```
Expect: m.type = evt ∨ m.type = dtc              Expect: m.type = rtd
 1  def ROUTE(m, p_t)                            10  def FORWD(m, p_t)
 2    p_t ! ⟨rtd, self(), m⟩                     11    p_t ! m
 3  end def                                      12  end def

 4  def TRACER(σ, v, p_s, p_t)                   13  def DETACH(p_s, p_t)
      # New (child) tracer state σ' is initialised with an   14    p'_t ← self()
      # empty routing map Π; instrumentation map σ.Φ is      15    PREEMPT(p_s, p'_t)
      # copied to σ', and the tracer-processes map Γ is      16    p_t ! ⟨dtc, p'_t, p_s⟩
      # initialised with the (first) process being traced, p_s  17  end def
 5    σ' ← ⟨Π ← ∅, σ.Φ, Γ ← {⟨p_s, •⟩}⟩
 6    DETACH(p_s, p_t)                           18  def TRYGC(σ, p_a)
 7    p_a ← fork(v) executable monitor           19    if σ.Γ = ∅ ∧ σ.Π = ∅ then
      # Tracer started in • mode to process routed trace  20      Signal analyser p_a to terminate
      # events first                              21      Terminate tracer
 8    LOOP_•(σ', p_a)                            22    end if
 9  end def                                      23  end def
```

Lst. 4. Operations used by the (∘) and priority (•) tracer loops

## A FURTHER OUTLINE INSTRUMENTATION DETAILS

Our message routing and forwarding operations described in sec. 2 enable tracers to implement hop-by-hop routing; these are given in lst. 4. The function self() on line 2 returns the PID of the calling process. Lst. 4 includes the TRACER function that is forked in lst. 1 to execute the core tracer logic of lsts. 2 and 3. DETACH is used to signal to the router tracer $p_t$ that the system process $p_s$ is being tracer by a new tracer, $p'_t$. Prior to issuing the message, detach invokes PREEMPT so that $p'_t$ takes over the tracing of system process $p_s$. TRYGC determines whether a tracer can be safely terminated. For the case of EA of fig. 1a, TRYGC also signals the analyser to terminate. The analyser terminates asynchronously so that it can process potential trace events it might still have in its message queue.

*Starting the system.* START in lst. 5 launches the SuS and monitoring system in tandem. The operation accepts the code signature $g$, as the entry point of the SuS, together with the instrumentation map, $\Phi$. As a safeguard that prevents the initial loss of trace events, the SuS is launched in a paused state (line 2) to permit the root tracer to start tracing the top-level system process. ROOT resumes the system (8), and begins its trace inspection in *direct* mode, as shown on line 10.

*Tracing.* The tracing mechanism is defined by the operations TRACE, CLEAR and PREEMPT listed in lst. 6.

## B INLINE INSTRUMENTATION

Our RV tool instruments monitors into the target system via *code injection* by manipulating the program abstract syntax tree (AST). Fig. 11 outlines how this process is carried out. In step ①, the Erlang source code of the system is parsed into the corresponding AST, step ②. The Erlang compilation process contains a *parse transform phase* [24], step ③, that provides a hook to enable the AST to be post-processed. The custom-built weaver leverages this mechanism in step

```
 1  def START(g, Φ)                              6  def ROOT(p_s, Φ)
      # Pausing allows root tracer to be set      7    TRACE(p_s, self())
      # up; no initial message loss               8    Resume system p_s
 2    p_s ← fork(g) in paused mode                9    σ ← ⟨Π ← ∅, Φ, Γ ← {⟨p_s, ∘⟩}⟩
 3    p_t ← fork(ROOT(p_s, Φ))                     # Root tracer has no monitor
 4    return ⟨p_s, p_t⟩                          10    LOOP_∘(σ, ⊥)
 5  end def                                      11  end def
```

Lst. 5. System starting operation and root tracer

A Choreographed Outline Instrumentation Algorithm for Asynchronous Components

```
 1  def Trace(p_s, p_t)                              11  def Clear(p_s, p_t)
 2    if p_s is not traced then                      12    if p_s is traced then
 3      Set the tracer for p_s to p_t; p_t will trace new    13      Clear the tracer p_t for p_s; p_t still traces
        descendant processes p_{s_1}, p_{s_2},... forked             the descendant processes p_{s_1}, p_{s_2},... of p_s
        by p_s automatically (assumption A_9)         14      Block until the trace events for p_s that are in
 4      while p_s's tracer is set do                           transit are delivered to p_t
 5        s ← read next event for p_s from           15    end if
            trace event source                       16  end def
 6        e ← encode s as a message                  Expect: p_s's tracer is set
 7        p_t ! e                                    17  def Preempt(p_s, p_t)
 8      end while                                    18    p'_t ← p_s's tracer
 9    end if                                         19    Clear(p_s, p'_t)
10  end def                                          20    Trace(p_s, p_t)
                                                     21  end def
```

Lst. 6. Tracing operations offered by the tracing mechanism

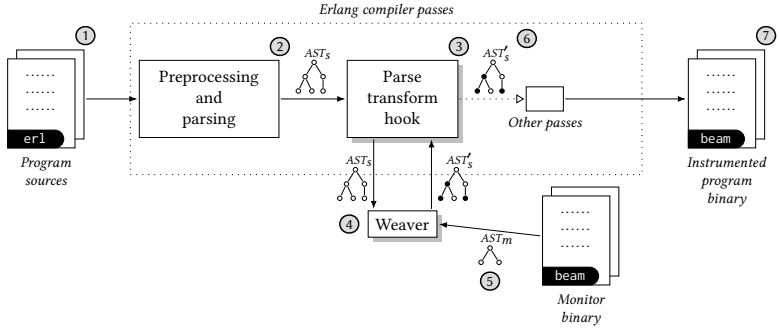

Fig. 11. Instrumentation pipeline for inline monitors

④, to embed into the program AST the AST of the synthesised monitor, step ⑤. The weaver performs two types of transformations:

  (i) *Monitor bootstrapping.* The function encoding the monitor is stored in the process dictionary (a process-local key-value store) of the monitored process.
 (ii) *Instrumentation points.* The program AST is instrumented with calls at the points of interest: these calls constitute the extraction and analysis of system trace events by the monitor code.

The instrumented calls in transformation *(ii)* retrieve the monitor function stored in the process dictionary in transformation *(i)*, and apply it to the system trace event. This function application returns the monitor continuation function that is used to replace the current monitor in the process dictionary. The two-step weaving procedure produces the instrumented program AST in step ⑥, that can be subsequently compiled by the Erlang compiler into the system binary, step ⑦.

## C  ADDITIONAL EXPERIMENT RESULTS

*Total memory consumed.* Fig. 12 shows the total memory consumed (left) and sampled memory (right) during the experiment runs conducted under Steady, Pulse and Burst loads for the case study with $n = 100$k workers. Note that unlike in figs. 9 and 10, the $y$-axis is labelled in GB. The total memory consumed (left) in fig. 12 corresponds to the *area*

| | Steady | | | | Pulse | | | | Burst | | | |
|---|---|---|---|---|---|---|---|---|---|---|---|---|
| | Master | | Per worker | | Master | | Per worker | | Master | | Per worker | |
| | EA | IA | EA | IA | EA | IA | EA | IA | EA | IA | EA | IA |
| **Scheduler utilisation** | 20.4 | 10.1 | 0.001 | 0.001 | 21.4 | 14.4 | 0.001 | 0.000 | 23.2 | 15.4 | 0.001 | 0.000 |
| **Memory consumption** | 1.4 | 0.7 | 0.000 | 0.000 | 2.0 | 1.5 | 0.000 | 0.000 | 1.4 | 0.9 | 0.000 | 0.000 |
| **RT** | 194.6 | 134.4 | 0.001 | 0.001 | 200.5 | 185.4 | 0.001 | 0.001 | 87.2 | 72.7 | 0.000 | 0.000 |
| **Execution duration** | 74.9 | 61.1 | 0.000 | 0.000 | 79.2 | 73.7 | 0.000 | 0.000 | 60.7 | 50.3 | 0.000 | 0.000 |

Tbl. 4. Percentage *estimated* runtime overhead per worker at the *maximum* load of 100k workers under the sampled memory plots (right). Our sampled memory plots reflect the shapes of the loads applied, although these extend for a longer duration that goes beyond the original loading time of $t = 100$s (see fig. 6).

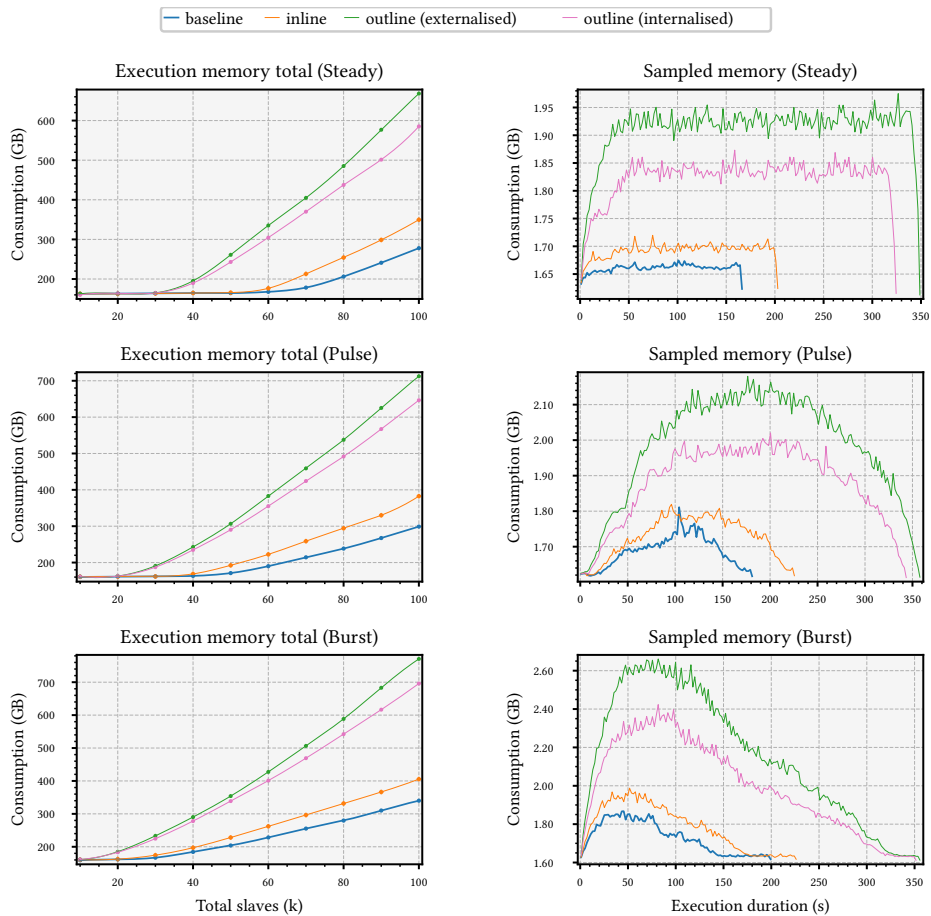

Fig. 12. Total and sampled memory consumption for benchmark configured with high load of 100 k workers

A Choreographed Outline Instrumentation Algorithm for Asynchronous Components

*Overhead on master process only.* Figs. 15 and 16 show the mean runtime overhead on the master process for the benchmark set up with 100 k workers. The combined overhead for the set-up for the master and worker processes is plotted for reference.

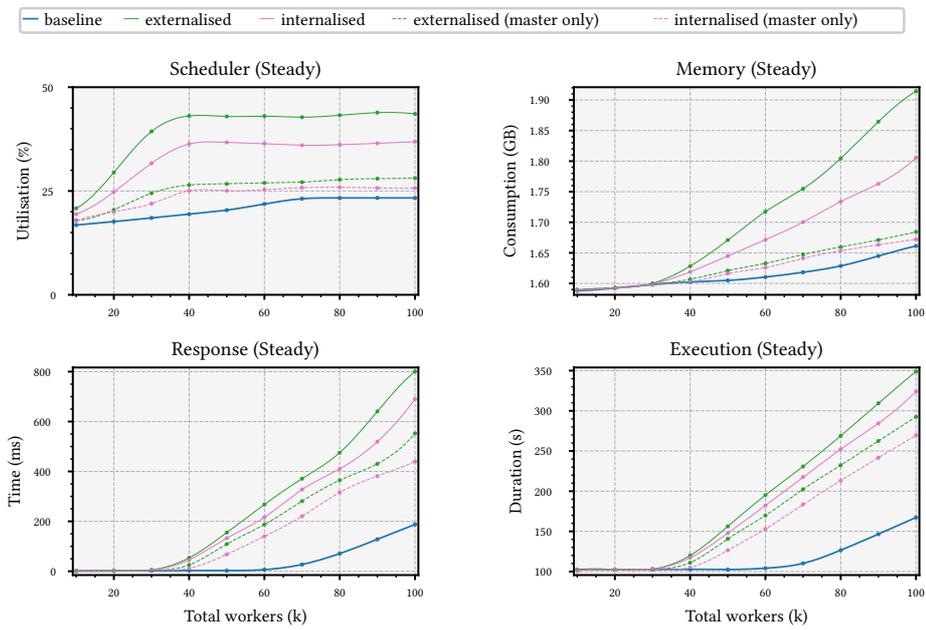

Fig. 13. Mean runtime overhead on master *only* for benchmark configured with high load (100 k workers)

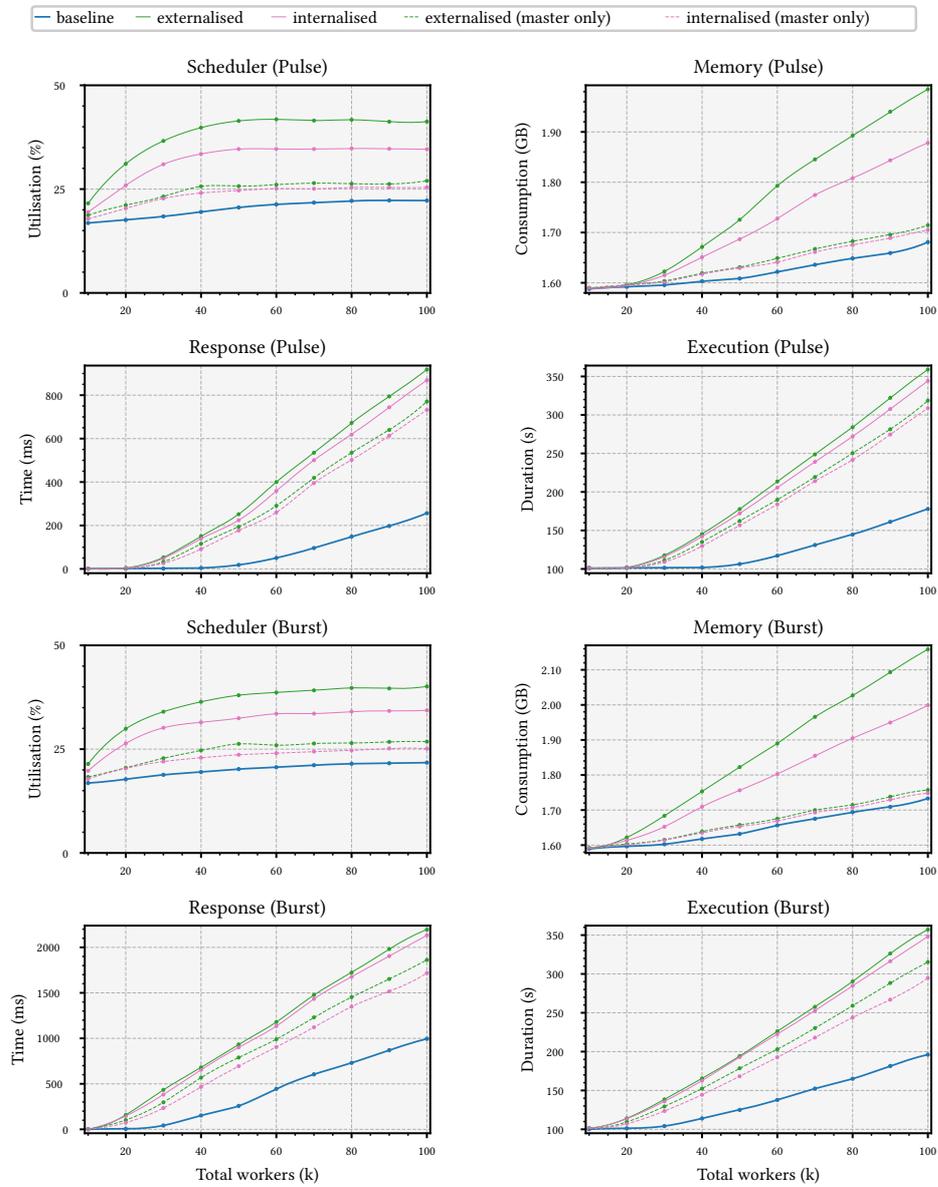

Fig. 14. Mean runtime overhead on master *only* for benchmark configured with high load (100 k workers, cont.)

A Choreographed Outline Instrumentation Algorithm for Asynchronous Components

*Estimated overhead per worker.* Figs. 15 and 16 show the estimated overhead per worker for the benchmark set up with 100 k workers.

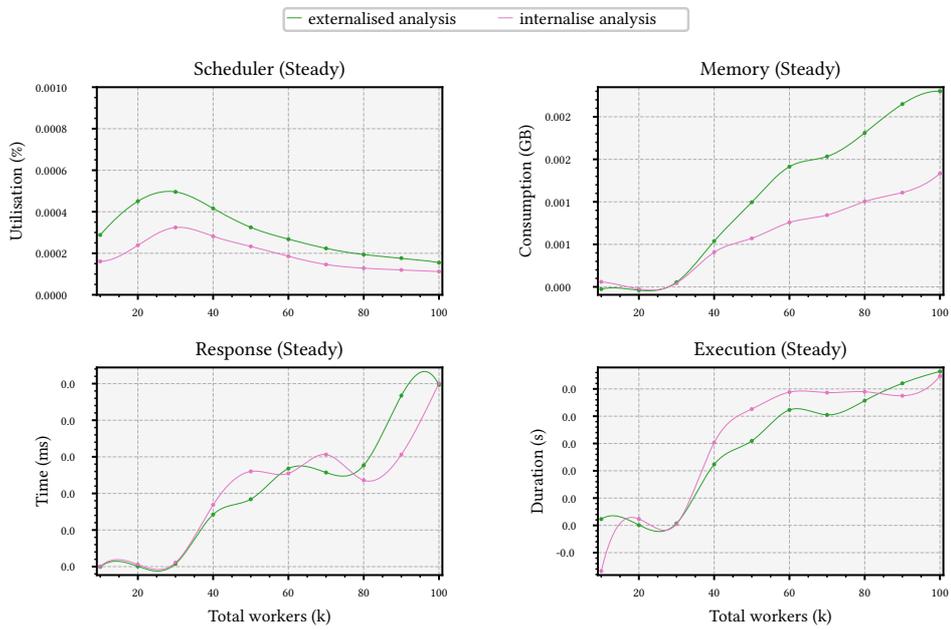

Fig. 15. Mean estimated runtime overhead per worker for benchmark configured with high load (100 k workers)

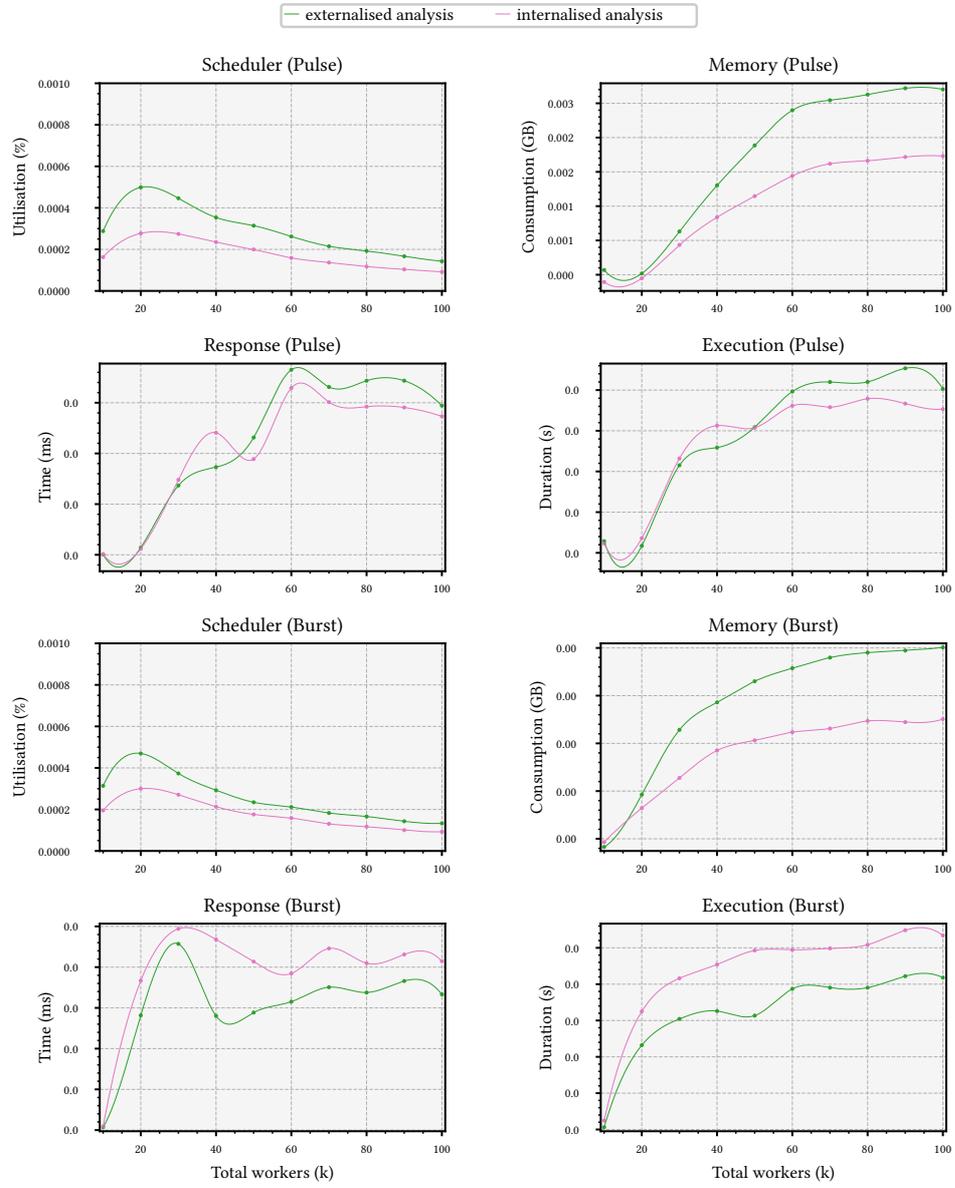

Fig. 16. Mean estimated runtime overhead per worker for benchmark configured with high load (100 k workers, cont.)

A Choreographed Outline Instrumentation Algorithm for Asynchronous Components

*Fitted data plots.* Figs. 17 and 18, and figs. 19 and 20 show the mean runtime overhead for the benchmark set up with 10 k and 100 k workers respectively. These plots, which correspond to figs. 7–10, have been fitted with linear, quadratic and cubic polynomials where the $R^2$ is above 0.96.

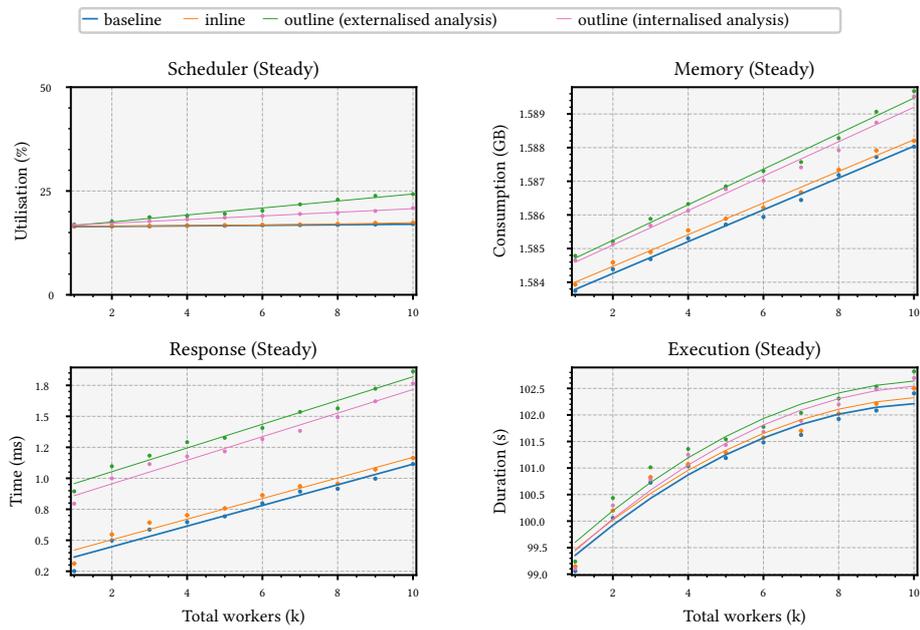

Fig. 17. Mean runtime overhead for benchmark configured with moderate load (10 k workers)

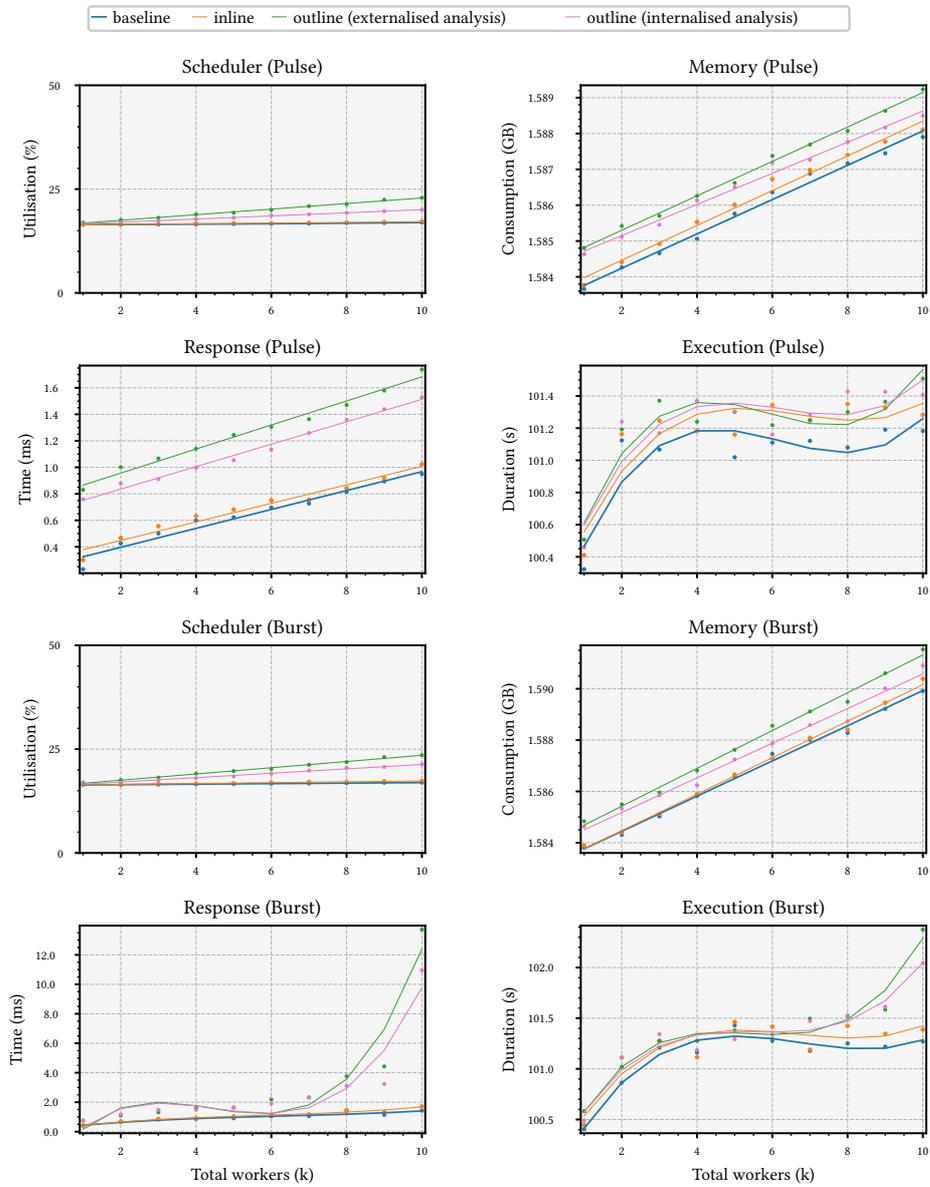

Fig. 18. Mean runtime overhead for benchmark configured with moderate load (10 k workers, cont.)

A Choreographed Outline Instrumentation Algorithm for Asynchronous Components

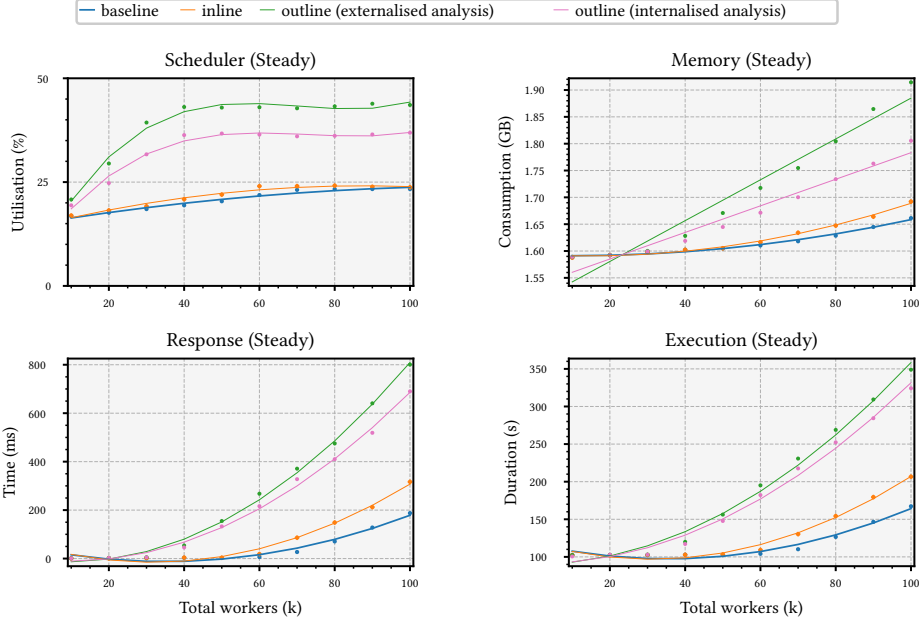

Fig. 19. Mean runtime overhead for benchmark configured with high load (100 k workers)

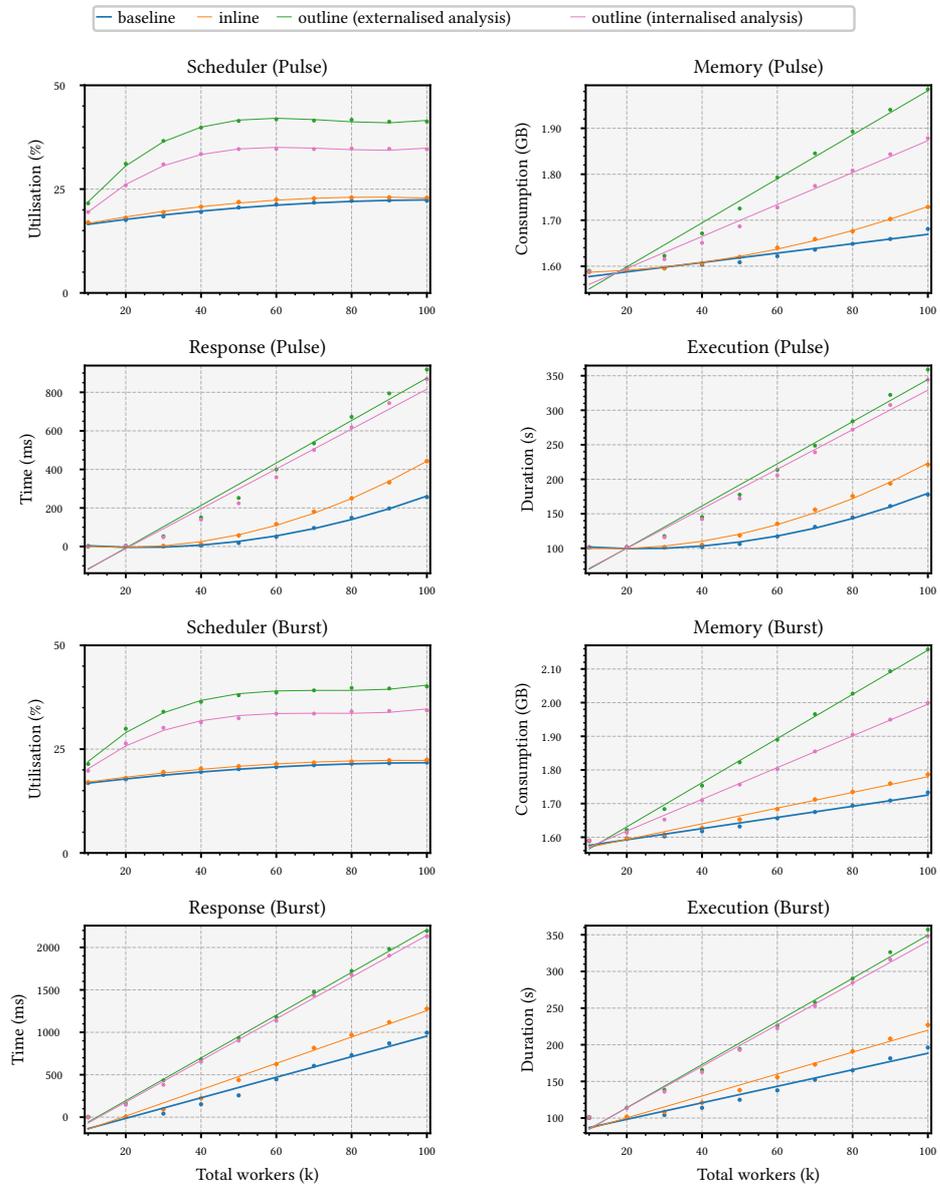

Fig. 20. Mean runtime overhead for benchmark configured with high load (100 k workers, cont.)